\documentclass[twocolumn]{aastex63}
\usepackage{natbib}
\usepackage{amsmath}
\received{2025 Mars 17}
\revised{2025 April 29}
\accepted{2025 May 2}

\newcommand\stf{star formation }
\newcommand\sfe{star formation efficiency }
\newcommand\sfing {star-forming }

\newcommand\gc{globular cluster }

\newcommand\gcs{globular clusters }

\newcommand\Ms{M_{\odot}}

\shorttitle{The ($m_{init}$,$F_{1P}$) distribution of Galactic Globular Clusters}
\shortauthors{Parmentier}

\begin{document}

\title{Cracking the relation between mass and 1P-star fraction of globular clusters: \\
III.~Initial distributions of in-situ and ex-situ clusters}

\correspondingauthor{Genevi\`eve Parmentier}
\email{gparm@ari.uni-heidelberg.de}

\author[0000-0002-2152-4298 ]{Genevi\`eve Parmentier}
\affiliation{Astronomisches Rechen-Institut, Zentrum f\"ur Astronomie der Universit\"at Heidelberg, M\"onchhofstr. 12-14, D-69120 Heidelberg, Germany}

\begin{abstract} 
Galactic globular clusters consist of two main stellar populations, the pristine (1P) and  polluted (2P) stars.  The fraction of 1P stars in clusters, $F_{1P}$, is a decreasing function of the cluster present-day mass, $m_{prst}$.  The information about cluster formation it contains has yet to be unlocked. 

Paper~I demonstrated that the observed distribution $(m_{prst},F_{1P})$ of Galactic globular clusters can result from a pristine-star fraction that is inversely proportional to their birth mass, $m_{ecl}$.  This relation was then calibrated with a fixed stellar mass threshold for 2P-star formation, $m_{th}$, i.e., $F_{1P}=m_{th}/m_{ecl}$.    

We now estimate the masses $m_{init}$ of Galactic globular clusters as they start their long-term gas-free evolution in the Galaxy and we map their behavior in the $(m_{init},F_{1P})$ space.  Several dissolution time-scales are tested (with and without primordial mass segregation), each yielding its own initial cluster distribution $(m_{init},F_{1P})$.  

The $(m_{init},F_{1P})$ distributions are mapped according to cluster origin, with the emphasis on the Disk, Low-Energy and Gaia-Enceladus cluster groups of \citet{mass19}.
All three initial distributions $(m_{init},F_{1P})$ are more compact than their present-day  counterparts since dynamical evolution scatters clusters in the $F_{1P}$ versus cluster-mass space.  The Disk initial distribution is the tightest one and potential reasons for this are discussed.  Its power-law representation allows us to generalize the initial mass threshold of Paper~I and prompts us to represent the cluster $({\rm mass},F_{1P})$ distribution in a log-log space.  No evidence is found suggesting that, initially, the pristine-star fraction of globular clusters depends on their metallicity on top of their mass.  

\end{abstract}


\section{Introduction}\label{sec:intro}

The puzzle of the multiple populations in Galactic \gcs and compact massive clusters in general remains unsolved.  Four decades ago already, stellar abundance analyzes revealed that Galactic \gcs are chemically inhomogeneous with respect to light elements \citep[see ][for an early comprehensive review]{kraft94}.  More recent spectroscopic analyzes have confirmed significant abundance spreads in He, C, N, O, Na and, in some clusters, Mg and Al \citep[e.g.][]{briley96,cannon98,car09b,smo11,muccia12,lardo13,carr19,mesza20,schia24}.  

Broadly speaking, Galactic \gcs consist of two main populations.  The first-population (1P or pristine) stars present a field-like chemical composition, while the second-population (2P or polluted) stars are carbon- and oxygen-poorer, helium-, nitrogen- and sodium-richer.  

Spectroscopic studies demonstrated early on that more massive clusters tend to host a greater fraction of polluted stars.   
Brighter \gcs present a higher number ratio of CN-strong to CN-weak stars (Fig.~2 in  \citealt{smith02} and Fig.~11 in \citealt{kayser08}) and more extended Na-O and Mg-Al anticorrelations (as measured by their interquartile range; Fig.~12 in \citealt{carretta06} and Fig.~16 in \citealt{car10a}; see also Fig.~5 in \citealt{panci17}).  With the brightness of \gcs a proxy to their present-day mass, these results show that the observed chemical inhomogeneities are stronger in more massive clusters.   

The Chromosome Map, a photometric tool that exploits the high sensitivity of stellar ultraviolet colours to CNO abundances \citep{mil15}, has increased the census of clusters with a measured pristine-star fraction, $F_{1P}$ \citep[e.g.][]{mil17,dondo21}, confirming early spectroscopic findings.  While the most massive clusters consist mostly of polluted stars ($F_{1P}\simeq0.2$), single-population clusters ($F_{1P}=1$) are all low-mass clusters.  That is, $F_{1P}$ decreases with the cluster present-day mass, $m_{prst}$, albeit with a significant scatter.
When considering clusters of similar masses, the pristine-star fraction is on average higher in more remote clusters \citep{zen19}.  Equivalently, at a given pristine-star fraction, more remote clusters are on average  more massive, as expected from the slower cluster evaporation at larger Galactocentric distances (assuming that all clusters share the same initial relation between mass and pristine-star fraction).  

\citet{par24a} demonstrates that the distribution of Galactic \gcs in the $(m_{prst},F_{1P})$ space can result from a pristine-star fraction that is inversely proportional to their birth mass, $m_{ecl}$.  Here,  $m_{ecl}$ is the stellar mass of newly formed clusters, as their massive-star activity is about to expel their residual embedding gas ("ecl" stands for "embedded cluster"). The relation $F_{1P}=m_{th}/m_{ecl}$ is calibrated with a fixed stellar-mass threshold for 2P-star formation, $m_{th}$ \citep[][Eq.~1]{par24a}.  In their model, clusters are polluted completely and instantaneously once their developing stellar mass has reached the threshold $m_{th}$ (i.e., in forming clusters, the mass in 1P stars does not grow beyond $m_{th}$).  The pristine-star fraction $F_{1P}$ is considered constant as clusters evolve, equivalently 1P and 2P stars are assumed to be well-mixed early on in the cluster history, as observed for some dynamically young \gcs \citep[][their Fig.~15]{leit23}.  

Evolving the relation $F_{1P}=m_{th}/m_{ecl}$ up to the age of 12\,Gyr yields a sequence of tracks that range from short to large Galactocentric distances, thereby covering the $(m_{prst},F_{1P})$ distribution of clusters \citep[Figs~7-8 in][]{par24a}.  The segregation between inner and outer clusters noticed by \citet{zen19} follows naturally since inner clusters endure greater mass losses than remote clusters.  The limits of the distribution on its top-right and bottom left sides appear to result from, respectively, the scarcity of clusters in the outer Galactic regions and the scarcity of massive clusters due to dynamical friction and size-of-sample effect \citep[for a comprehensive summary, see Sec.~2.1 in][]{par24b}.  
The model also accounts for the tight correlation of the pristine-star fraction with the mass in polluted stars, and for its poor correlation with the mass in pristine stars (Fig.~7 in \citealt{mil20}, Fig.~4 in \citealt{par24b}). 

To calibrate the mass threshold $m_{th}$ for 2P-star formation remains a task fraught with uncertainties, given how fragmentary our knowledge of multiple-population cluster formation is.  \citet{par24a} bypassed the difficulty by calibrating the product $m_{th}F_{bound}^{VR}F_{StEv}$, namely, the threshold $m_{th}$ reduced by, firstly, violent relaxation \citep[i.e., the escape of stars that follows residual \sfing gas expulsion; e.g.][]{hills80,geyer01,bk07}, then by stellar-evolution mass losses.  Here, $F_{bound}^{VR}$ and $F_{StEv}$ are the mass fractions retained by clusters after violent relaxation and stellar-evolution mass losses.  In what follows, we refer to the initial mass $m_{init}=F_{bound}^{VR}m_{ecl}$ as the cluster mass at the end of violent relaxation, equivalently at the beginning of their long-term gas-free evolution in the Galactic potential (aka secular evolution).  The track $F_{1P}=m_{th}/m_{ecl}$ shifted down in mass by the factor $F_{bound}^{VR}F_{StEv}$ corresponds therefore to clusters that lose none of their stars during  secular evolution (although they have lost a fraction $1-F_{bound}^{VR}$ of their stellar mass during violent relaxation).  Such clusters must be massive and remote, two factors slowing down cluster evaporation \citep[e.g.~][]{meyheg97}.  NGC2419 is such a cluster \citep{baum19}.  

To anchor NGC~2419 on the shifted track $F_{1P}(F_{StEv}F_{bound}^{VR}m_{ecl})$ allows one, for a given $F_{StEv}$ value, to estimate the mass threshold $m_{th,init}=F_{bound}^{VR} m_{th}$ that clusters must have to {\it host} 2P stars at secular-evolution onset, even though the bound fraction $F_{bound}^{VR}$ and the mass threshold $m_{th}$ for 2P-star {\it formation} remain ill-constrained \citep[Fig.~5 in][]{par24a}. 

But despite its large mass and remote location, NGC~2419 may not be the ideal calibrator.  A member of the Sagittarius dwarf galaxy \citep{mass19,forbes20}, it is a latecomer in the Galactic gravitational potential \citep[first pericentric passage into the Milky Way 5.7\,Gyr ago,][]{ruiz20}.  The Mg-abundance distribution of its giant stars is also anomalous, stretching down to [Mg/Fe] $\simeq-1$dex (\citealt{muccia12}; see also Fig.~6 in \citet{panci17} for its mean [Mg/Fe] and [Mg/$\alpha$] abundance ratios).  NGC~2419 looks like an intruder among most other Galactic globular clusters.  

Furthermore, the approach adopted in \citealt{par24a} and \citealt{par24b} (hereafter Paper~I and Paper~II) is a forward-looking approach.  That is, it assumes a given initial track $F_{1P}(m_{init})$, the same for all clusters, and evolves it up to the age of 12\,Gyr.  In the present paper, we adopt instead a backward-looking approach, in which the initial masses  $m_{init}=F_{bound}^{VR} m_{ecl}$ of Galactic \gcs are estimated based on their present-day mass and orbit.  Once cluster present-day masses have been corrected for the effects of stellar and secular evolutions, the cluster $(m_{init},F_{1P})$ distribution can be investigated.
   
\begin{figure*}
\includegraphics[width=0.49\textwidth, trim = 0 10 0 0]{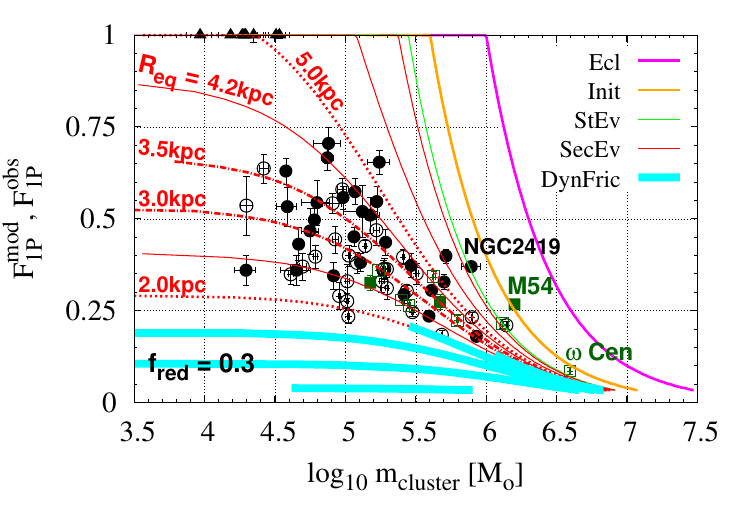}
\includegraphics[width=0.49\textwidth, trim = 0 10 0 0]{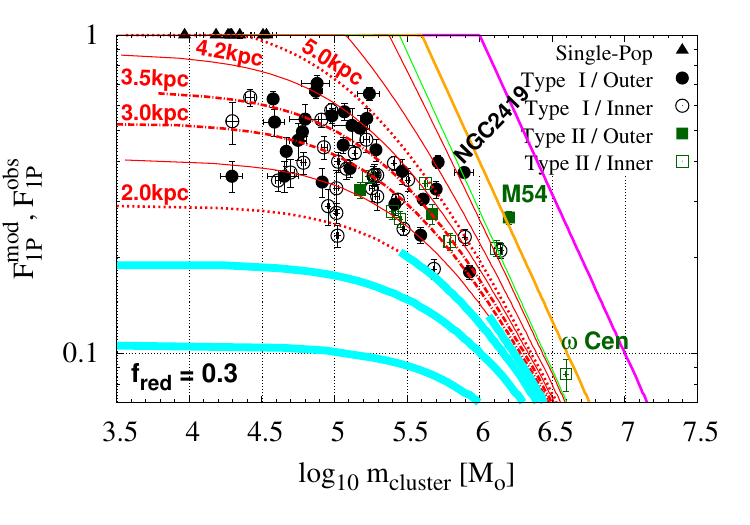}  \\
\caption{Left panel: Relation between cluster mass $m_{cluster}$ and pristine-star fraction $F_{1P}$.  Symbols depict present-day Galactic globular clusters, with triangles, circles and squares standing for, respectively, single-population clusters, Type~I and Type~II multiple-population clusters.  Open and plain symbols represent inner ($R_{eq}\leq 3.1$\,kpc) and outer ($R_{eq}>3.1$\,kpc) clusters.  Lines are the model tracks from Fig.~7 in Paper~I (see text for details).  Right panel: Same as left panel with $F_{1P}$ on a logarithmic scale.  $\omega$~Cen stands here even more apart from the other globular clusters.}
\label{fig:loglog} 
\end{figure*} 

This approach allows us (1)~to generalize the fixed mass threshold for 2P-star formation introduced in Paper~I, (2)~to uncover the dispersion in the $(m_{init},F_{1P})$ space (not doable when assuming a one-to-one relation $F_{1P}(m_{init})$), and (3)~to compare the initial cluster distributions as a function of cluster origin.  For now, we keep assuming a constant pristine-star fraction.  
A forthcoming paper will present models accounting for the preferential loss of pristine or polluted stars, hence for a time-varying $F_{1P}$. \\

The present paper is organized as follows.  
Section~\ref{sec:loglog} substitutes a log-log representation to the usual logarithmic-linear depiction of the cluster $({\rm mass},F_{1P})$ distribution.  Cluster initial masses are estimated in Sec.~\ref{sec:minit}.  Section~\ref{sec:origin} maps the initial relations $F_{1P}(m_{init})$ according to cluster origin and Sec.~\ref{sec:hdisp} discusses their respective dispersions.  Whether the pristine-star fraction partly depends on metallicity is discussed in Sec.~\ref{sec:FeH}.  Section~\ref{sec:phsp} discusses the uncertainties affecting the pristine-star fractions derived from photometry.  Section~\ref{sec:ninst} moves beyond the fixed stellar-mass threshold of Paper~I.  We conclude in Sec.~\ref{sec:conclu}.

\section{A log-log plot for the $({\rm mass}, F_{1P})$ space} \label{sec:loglog}

The relation between cluster mass and pristine-star fraction is traditionally represented with $F_{1P}$ on a linear scale and the cluster mass on a logarithmic scale.  With cluster masses covering  several orders of magnitude and pristine-star fractions stretching over a single one \citep[the lowest pristine-star fraction known is that of $\omega$~Cen, $F_{1P} \simeq 0.09$;][]{mil17,dondo21}, this may seem a logical choice.  But the power-law relation $F_{1P}=m_{th,init}/m_{init}=m_{th}/m_{ecl}$ adopted in Paper~I prompts us to move to a log-log space.  

Both perspectives are shown in Fig.~\ref{fig:loglog}.  Symbols depict our Galatic \gc sample, as described in Sec.~2.2 of Paper~II.  Lines are the model tracks from Paper~I.  The magenta and orange tracks show the relations $F_{1P}(m_{ecl})$ at gas-expulsion onset and $F_{1P}(m_{init})$ at secular-evolution onset, with $m_{th}=10^6\,\Ms$ and $F_{bound}^{VR}=0.4$.  The green track shows the relation after stellar-evolution mass losses, assuming $F_{StEv}=0.70$.  12\,Gyr-old tracks are shown in red for cluster orbital equivalent radii $R_{eq}=(1-e)D_{apo}=30.0,8.0,5.0,4.2,3.5,3.0,2.5,2.0,1.5,1.0{\rm ~~and~~}0.5$\,kpc (the 30\,kpc-track is the one closest to the green track).  $e$ and $D_{apo}$ are the cluster-orbit eccentricity and apocentric distance.  Short-dashed lines mark the tracks for $R_{eq}=$2.0 and 5.0\,kpc, and dashed-dotted lines for $R_{eq}=$3.0 and 3.5\,kpc.  The equivalent radius $R_{eq}$ was introduced in Paper~I as it scales linearly with the cluster dissolution time-scale $t_{diss}$ \citep[Eq.~10 in][]{bm03}.  The smaller $R_{eq}$, the closer to the Galactic center clusters venture and the faster they dissolve.  While the red tracks stick to the post-stellar-evolution mass-losses (green) track at high mass, their low-mass sections stretch out leftward at short $R_{eq}$, highlighting the final dissolution of clusters in a strong tidal field.  We recall that a constant pristine-star fraction $F_{1P}$ is assumed (i.e., model clusters move leftward in Fig.~\ref{fig:loglog}).  Thick cyan lines mark the region of the diagram impacted by dynamical friction, assuming that clusters are initially on circular orbits of radius $R_{eq}$.  Sec.~2.1 of Paper~II provides a more detailed summary.  The parameter $f_{red}$ is re-explained in Sec.~\ref{sec:minit}.   

In Papers~I and II, the massive and remote cluster NGC~2419 is used to "anchor" the green and orange tracks.  With $F_{StEv}=0.7$, this results in a mass threshold $m_{th,init}\simeq 4\cdot10^5\,\Ms$ for clusters to host multiple populations at secular-evolution onset (intersection between orange track and top $x$-axis).     

In the log-log space (right panel), the power-law relations $F_{1P}=m_{th}/m_{ecl}$ and $F_{1P}=m_{th,init}/m_{init}$ of Paper~I (magenta and orange solid lines) become  straigth lines of slope $-1$.  
In what follows, we keep refereing to the $({\rm mass},F_{1P})$ space, although our figures will show $\log_{10}(F_{1P})$ in dependence of $\log_{10}(m_{cluster})$.  Here, $m_{cluster}$ is the cluster mass at a specific evolutionary stage (i.e.~$m_{cluster}=m_{ecl}$ and $m_{cluster}=m_{init}$ for the orange and magenta tracks, respectively, and $m_{cluster}=m_{prst}$ for the red tracks).

\section{Cluster initial masses} \label{sec:minit}

To estimate the initial masses $m_{init}$ of clusters, we need their dissolution time-scales,  $t_{diss}$.  
Papers~I and II define it as $t_{diss} = f_{red} t_{diss}^{BM03}$, namely, the cluster dissolution time-scale of \citet{bm03} times a reduction factor $f_{red}<1$.  In the $N$-body modeling of \citet{bm03}, model clusters orbit a spherically-symmetric logarithmic potential of circular velocity $V_c=220\,km\cdot s^{-1}$, they obey a \citet{kro01} stellar Initial Mass Function (IMF) and are not primordially mass segregated.   The reduction factor $f_{red}$ accounts for the faster dissolution of clusters that are primordially mass segregated \citep{hag14} and/or whose IMF is top-heavy  \citep{hs20}.  The exact value of $f_{red}$ is unknown, however, as neither the degree of primordial mass segregation nor the IMF of old \gcs is well-constrained. 

Paper~I estimated the reduction factor at $f_{red}=0.3$ by forcing the track for $R_{eq}=3.1$\,kpc, our cluster-sample median equivalent radius, to part the cloud of data points in two roughly equal-size parts (see Fig.~\ref{fig:loglog}).  But this estimate is tied to Paper~I 's initial relation $F_{1P}=4\cdot10^5\,\Ms / m_{init}$.  In this contribution, we shall thus test different $f_{red}$ values, each resulting in a set of cluster initial mass estimates. \\

Our cluster initial mass estimates build on Eqs~10 and 12 of \citet{bm03}, which, combined with the reduction factor $f_{red}$, yield
\begin{equation}
m_{init}=\frac{m_{prst}}{F_{StEv}}+\frac{10^3\tau(V_c/220)\bar{m}_*^x}{\beta f_{red}R_{eq}} m_{init}^{1-x}\left[\ln\left(0.02\frac{m_{init}}{\bar{m}_x}\right)\right]^x
\label{eq:minit}
\end{equation}  
$m_{prst}$ and $\tau$ are the cluster present-day mass and age (in units of $\Ms$ and Gyr).  $R_{eq}$ is in units of kpc.  The parameters $x$ and $\beta$ determine the cluster initial King concentration, $W_0$ \citep[$W_0=5.0$: $(x,\beta)=(0.75,1.91)$; $W_0=7.0$: $(x,\beta)=(0.82,1.03)$;][]{bm03}. 
$\bar{m}_*$ is the mean stellar mass and $V_c$ is the circular velocity of the host-galaxy logarithmic potential, in units of $km \cdot s^{-1}$.  

With the initial mass $m_{init}$ on both its sides, Eq.~\ref{eq:minit} is solved iteratively.  Except for $f_{red}$, model parameters are as in Paper~I: $(x,\beta)=(0.75,1.91)$, $F_{StEv}=0.7$, $\bar{m}_*=0.55\Ms$, $V_c=220\,km\cdot s^{-1}$, $\tau=12$\,Gyr.  Cluster present-day masses and orbital data are taken from \citet{baum19} (version March2023, available at \url{https://people.smp.uq.edu.au/HolgerBaumgardt/globular/}). \\

\begin{figure}
\begin{center} 
\includegraphics[width=0.49\textwidth, trim = 0  20  0 0]{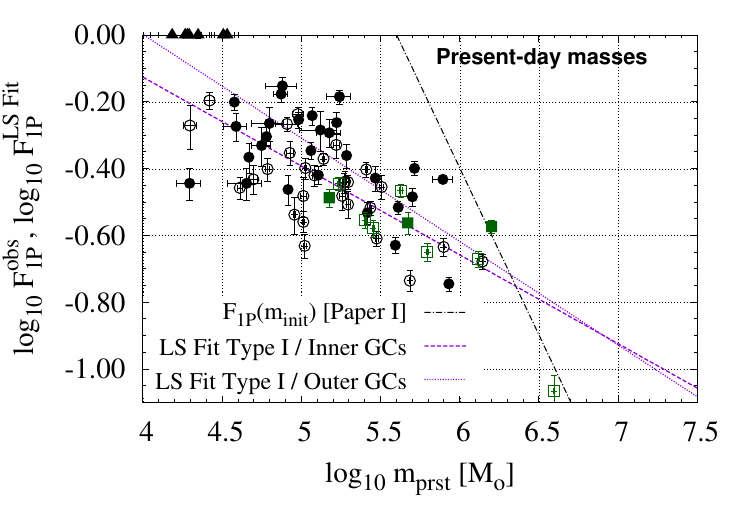} \\
\includegraphics[width=0.49\textwidth, trim = 0  65  0 0]{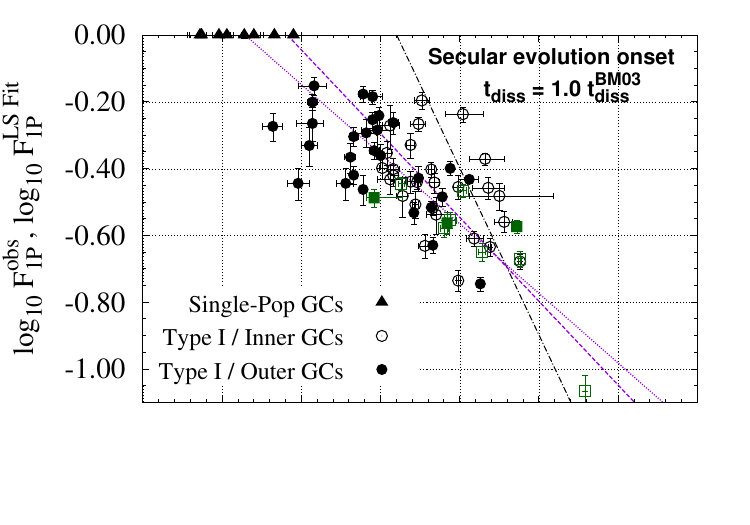} \\
\includegraphics[width=0.49\textwidth, trim = 0  00  0 0]{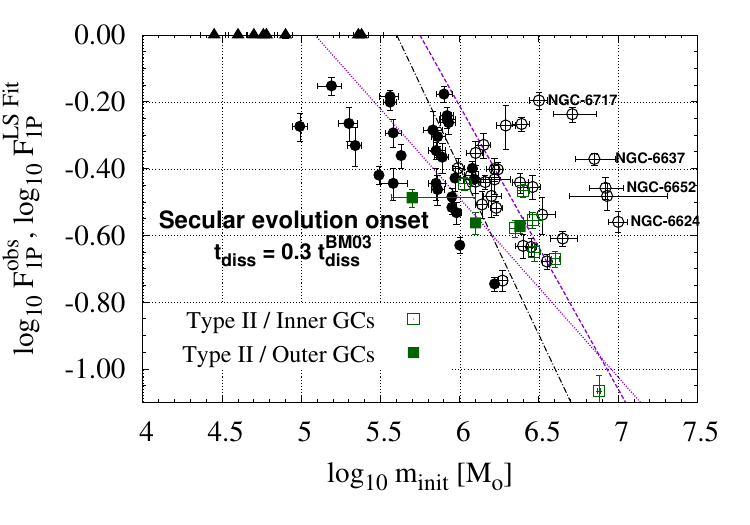} 
\caption{Top panel: Pristine-star fraction $F_{1P}$ versus present-day mass $m_{prst}$ of the  Galactic globular clusters of our sample.  Symbol coding as in Fig.~\ref{fig:loglog}.  
The violet dashed and dotted lines show the least-squares fits to the inner and outer multiple-population clusters of Type~I.  The dashed-dotted line is the initial track from Fig.~\ref{fig:loglog}.  Middle panel: Pristine-star fraction $F_{1P}$ versus \gc initial mass $m_{init}$ estimated from Eq.~\ref{eq:minit} with $f_{red}=1.0$ \citep[i.e., the cluster dissolution time-scale is that of][]{bm03}.  Bottom panel: same with $f_{red}=0.3$ (i.e., the cluster dissolution time-scale of \citealt{bm03} is reduced by a factor 0.3).  $F_{1P}$ is assumed to stay constant in time, i.e., data points move leftward from the bottom to the top panel.}
\label{fig:revobs} 
\end{center}
\end{figure}

The 1P-star fraction in clusters in dependence of their initial mass estimate $m_{init}$ is shown for $f_{red}=1.0$ and $0.3$ in the middle and bottom panels of Fig.~\ref{fig:revobs}.  $f_{red}=1.0$ corresponds to the cluster dissolution time-scale of \citet{bm03}, $f_{red}=0.3$ to the dissolution time-scale of a strongly primordially mass segregated cluster \citep{hag14}.  For the sake of comparison, the top panel shows $F_{1P}$ as a function of the cluster present-day mass.  Data points thus shift rightward from the top to the bottom panel.  The errors on $m_{init}$ are obtained by propagating the errors on the apo- and pericentric distances of cluster orbits, on the cluster present-day masss \citep[taken from][]{baum19} and ages ($\pm1$\,Gyr).  Symbol-coding is as in Fig.~\ref{fig:loglog}.   The dashed-dotted line, common to all panels, represents the initial track of Paper~I (solid orange track in Fig.~\ref{fig:loglog}).  

A dearth of low-mass inner clusters characterizes the middle and bottom panels.  For instance, when $f_{red}=1.0$, no inner cluster (open symbols) is initially less massive than $\simeq 3\cdot10^5\,\Ms$.  By contrast, the initial mass of outer clusters (plain symbols) ranges to $\lesssim 10^5\,\Ms$ (ignoring single-population clusters).  This highlights the incompleteness of our initial distributions.  We are indeed unable to recover the clusters that have dissolved over the past Hubble-time, an effect that is at its strongest for low-mass clusters on inner orbits.  

While all clusters move to higher masses from the top to the bottom panel, inner clusters experience larger mass corrections because of the stronger tidal field they endure.  They become more massive than most outer clusters when $f_{red}=0.3$, a configuration opposite to that today (compare top and bottom panels in Fig.~\ref{fig:revobs}).  An analogy could be made with an athletes race: when $f_{red}=0.3$, inner clusters set off from a starting line that is behind that of outer clusters (i.e., inner clusters are initially more massive), but, inner clusters being faster runners (i.e., they lose mass more quickly), they eventually overtake their contenders (i.e., their mass 12\,Gyr later is on average lower than that of outer clusters).

Interestingly, \citet{frxburn09} suggest that the inner halo formed the most massive globular clusters \citep[but see][]{car10a}.  They base their conclusion on inner-halo clusters hosting population II Cepheids and hotter horizontal-branch stars (their Fig.~2), which requires stronger helium enrichment.  A word about the terminology here: our inner clusters (not '{\it inner-halo}' clusters) also contain all bulge clusters (since they obey $R_{eq}=(1-e)D_{apo}\leq 3.1$\,kpc).  They thus cannot fully match the inner-halo group of \citet{frxburn09}.  Nevertheless, among the 25 inner-halo clusters of \citet{frxburn09}, 17 are in common with our data set and 14 are part of our inner-cluster group.  This represents $\gtrsim 40\,\%$ of the open symbols of our $(m_{cluster},F_{1P})$ plots.  Assuming that all clusters retain a similar fraction $F_{bound}^{VR}$ of their stars after gas expulsion (that is, inner clusters are more massive than outer clusters at birth already), the bottom panel of Fig.~\ref{fig:revobs} is reminiscent of \citet{frxburn09} 's suggestion that the inner halo formed more massive globular clusters than the outer halo and the disk (note that most disk clusters are part of our outer group; see Sec.~\ref{ssec:disk}).  That some inner clusters are initially more massive than the outer clusters could also explain why both populations partly overlap in the present-day space, as opposed to the neat order of the Papers~I and II model tracks (see Fig.~\ref{fig:loglog} and also Sec.~5 in Paper~I for additional factors).     


We caution that 4 of the 6 clusters that detach themselves from the rest of the group when $f_{red}=0.3$ (labelled rightward of their respective symbol in the bottom panel of Fig.~\ref{fig:revobs}) are bulge clusters \citep[Table~A.1 in][]{mass19} whose initial mass is overestimated due to Eq.~\ref{eq:minit} not accounting for dynamical friction (see also Sec.~\ref{sec:hdisp}).  Owing to their proximity to the Galactic center, bulge clusters are more affected by dynamical friction than any other cluster group and Sec.~\ref{sec:origin} will therefore not consider them.  

The dashed and dotted violet lines are the least-squares fits to the inner and outer clusters, respectively.  
Both fits steepen from the top to the bottom panels (as mass corrections are larger at low mass than at high mass), the steepening being stronger for the inner-cluster group than for the outer-cluster group (since inner clusters evolve faster; see also red tracks in Fig.~\ref{fig:loglog}).  These fits illustrate how the uncertain cluster dissolution time-scale impacts the mass threshold $m_{th,init}$ beyond which clusters host multiple populations at secular-evolution onset (recall that the mass threshold for 2P-star formation $m_{th}$ is higher by a factor $1/F_{bound}^{VR}$).  If $f_{red}=1.0$, we conclude that inner and outer clusters share fairly similar initial mass thresholds $m_{init,th}\simeq 6\cdot10^4\,\Ms$.  But if $f_{red}\simeq 0.3$, their respective threshold masses differ significantly on top of being higher.  

This is no safe approach, however, to try to reach conclusions based on the entire cluster sample.  The Galactic \gc system consists of distinct subsystems, from in-situ disk and bulge clusters, to ex-situ clusters accreted along with their natal dwarf galaxies.  

In the next section, we use the classification of \citet{mass19} to investigate how clusters behave in the $(m_{init},F_{1P})$ space as a function of $f_{red}$ and of their formation site (for alternative classifications of Galactic \gcs according to their origin, see \citealt{callingham22}, \citealt{horta20} and \citealt{forbes20}).

\section{The initial distribution $F_{1P}(m_{init})$ according to cluster origin}
\label{sec:origin}

\begin{figure}
\includegraphics[width=0.49\textwidth, trim = 0 20 0 0]{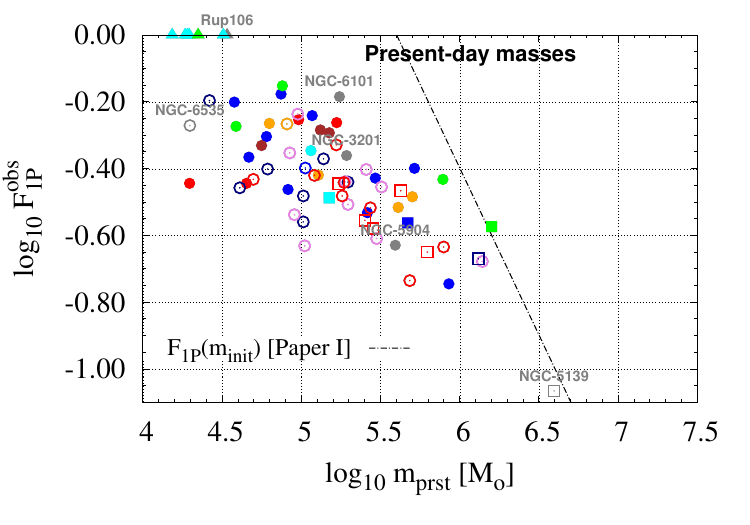}  \\
\includegraphics[width=0.49\textwidth, trim = 0 65 0 0]{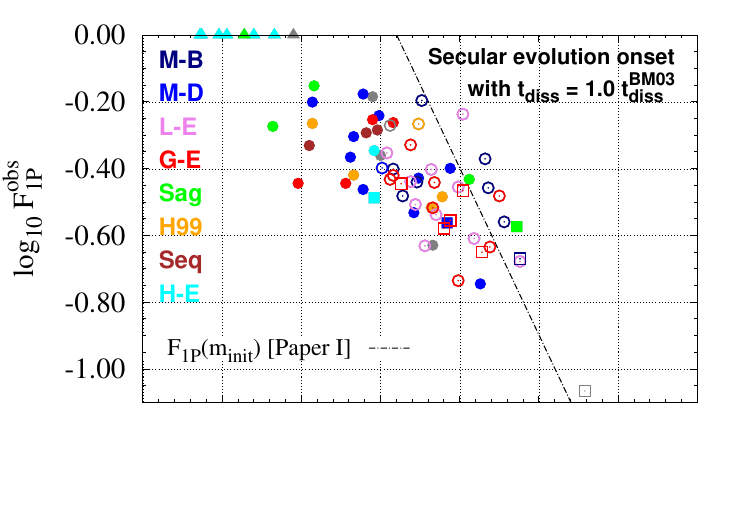}  \\
\includegraphics[width=0.49\textwidth, trim = 0  0 0 0]{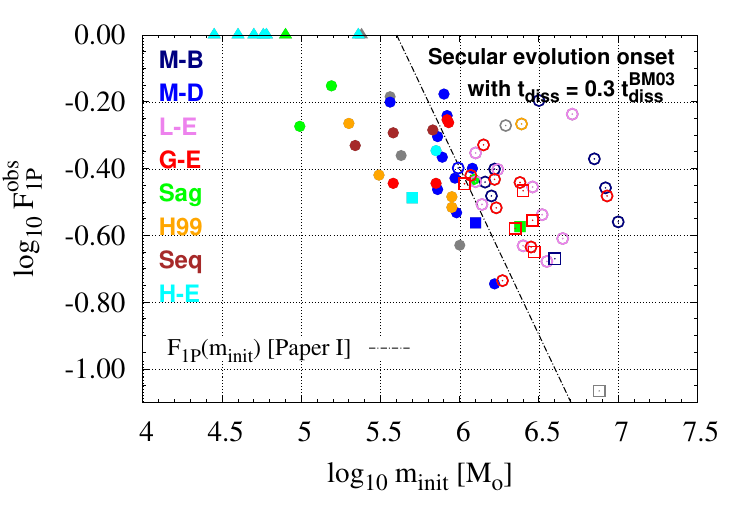}
\caption{Same as Fig.~\ref{fig:revobs} with the color coding marking the cluster origin.   Labels and color coding as in Table~A.1 and Fig.~2 of \citet{mass19} (see middle and bottom panels of this figure; 'M-B': bulge clusters; 'M-D': disk clusters; 'L-E': low-energy group; 'G-E': Gaia-Enceladus clusters; 'Sag': Sagittarius clusters; 'H99': Helmi Stream clusters; 'Seq': Sequoia clusters; 'H-E': high-energy clusters).  Clusters not unambiguously associated to a merger event by \citet{mass19} are left in gray and labelled in the top panel.  Symbol coding as in Fig.~\ref{fig:revobs} (circles: Type~I clusters, squares: Type~II clusters, triangles: single-population clusters; plain and open symbols: outer and inner clusters, respectively).   Error bars are not shown for the sake of clarity.    }  
\label{fig:origin} 
\end{figure} 

Building on the six-dimensional phase-space information that became available following the Gaia-mission second data release \citep{gaia18a}, \citet{mass19} identified the likely origin of 151 Galactic globular clusters, from in-situ bulge and disk clusters (labelled 'M-B' and 'M-D' in their Table~A.1) to accreted clusters.  Associated merger-event progenitors are the Gaia-Enceladus, Sagittarius, and Sequoia dwarf galaxies \citep[labelled 'G-E', 'Sag' and 'Seq' in Table A.1 of][]{mass19}, the progenitor of the Helmi Stream ('H99') and the host of a group of low-energy (i.e., tightly bound to the Galaxy) clusters ('L-E').  \citet{mass19} also identify a group of high-energy (i.e., loosely bound to the Galaxy) clusters of diverse origins ('H-E').  Figure~\ref{fig:origin} shows the cluster sample of our Fig.~\ref{fig:revobs} color coded for cluster origins (color coding as in Fig.~2 of \citealt{mass19}).  Clusters that are not unambiguously associated to a merger event are left in gray and labelled in the top panel.  For instance, $\omega$~Cen (NGC5139; $F_{1P}\simeq 0.09$) could be associated to either Gaia-Enceladus or Sequoia.  Three of the clusters lacking an unambiguous association in \citet{mass19} are linked to Sequoia (NGC3201 and NGC6101) and the Helmi Stream (NGC5904) by both \citet[][his appendix]{forbes20} and \citet[][their Table~A.1]{callingham22}.

We now consider the richest groups of clusters of our sample, namely, the disk (12 clusters), low-energy (10 clusters) and Gaia-Enceladus (17 clusters) groups.  In comparison, the Sequoia, Sagittarius and Helmi-stream groups contain 3, 5 and 5 clusters, respectively.  Table A.1 of \citet{mass19} contains more of these clusters, but only those with pristine-star fraction estimates are of interest here.  To increase, even if sligthly, the number of clusters of each group, Type~I and Type~II clusters are considered together.

\subsection{The disk clusters}
\label{ssec:disk}

\begin{figure}
\includegraphics[width=0.49\textwidth,trim = 0 20 0 0]{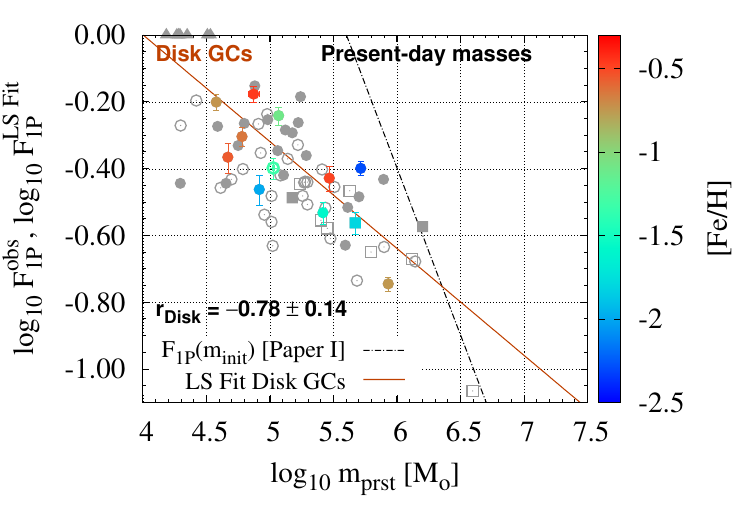} \\
\includegraphics[width=0.49\textwidth,trim = 0 60 0 0]{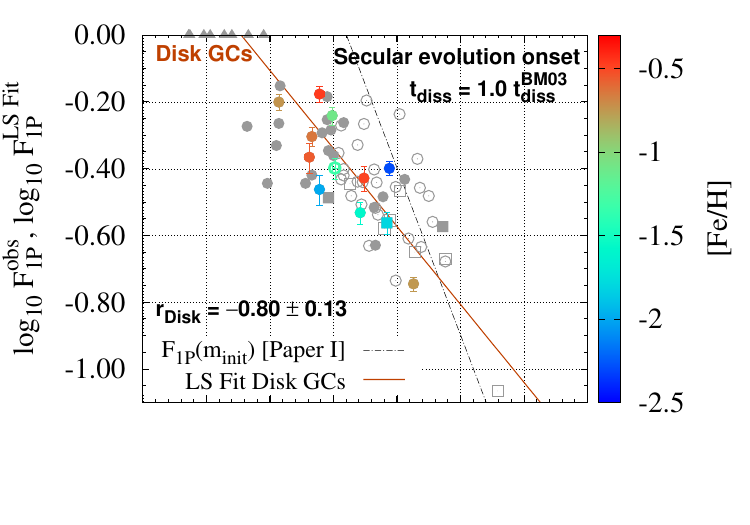} \\
\includegraphics[width=0.49\textwidth,trim = 0  0 0 0]{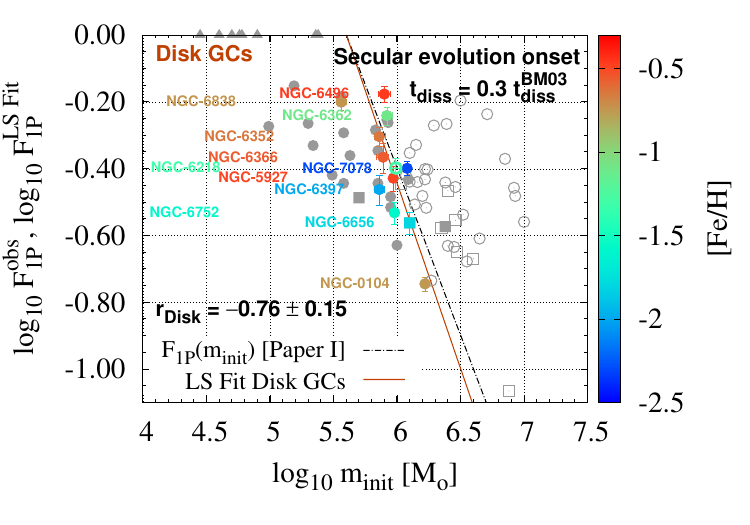}
\caption{Top panel: distribution of our globular cluster sample in the $(m_{prst},F_{1P})$ space with the twelve disk clusters of \citet{mass19} color-coded according to their  mean metallicity (see palette).  Each panel shows the disk-group correlation coeffficient $r_{Disk}$ and  least-square fit (solid dark-orange line).  The black dashed-dotted line is the initial relation from Paper~I.  Middle panel: same for the $(m_{init},F_{1P})$ space with  $f_{red}=1.0$.  Bottom panel: same for the $(m_{init},F_{1P})$ space with $f_{red}=0.3$.  The disk \gcs are labelled with the same color coding and at the same vertical location as their respective symbols.}  
\label{fig:disk} 
\end{figure}

The disk clusters \citep["M-D" in Table~A.1 of][]{mass19} are color-coded according to their mean metallicity in Fig.~\ref{fig:disk}, from $[Fe/H]=-2.29$ for NGC~7078 up to $[Fe/H]=-0.46$ for NGC~6496 (see palette).  All other clusters are left in gray.  The metallicities are taken from Table~1 in \citet{bai19} or, when not available there, from the McMaster catalog \citep[][version 2010]{har96}.  Disk clusters are labelled in the bottom panel with the same color coding and at the same vertical location as their corresponding symbol.  For the sake of clarity, only disk-cluster error bars are shown.  The dashed-dotted black line is the initial relation from Paper~I ($F_{1P}(m_{init})=4\cdot 10^5\,\Ms~m_{init}^{-1}$).       

Our disk-cluster sample contains one Type~II cluster (NGC~6656) and 11 Type~I clusters.  All but one are outer clusters (the lone inner cluster is NGC~6218 for which $R_{eq}=2.97\,{\rm kpc}<3.1$\,kpc).  That disk clusters are predominantly outer clusters stems from \citet{mass19} parting the bulge from the disk clusters at an apocentric distance $D_{apo}=3.5$\,kpc, slightly higher than the equivalent radius at which we part inner from outer clusters ($R_{eq}=(1-e)D_{apo}=3.1$\,kpc).  

\begin{figure}
\includegraphics[width=0.49\textwidth, trim = 0  0 0 0]{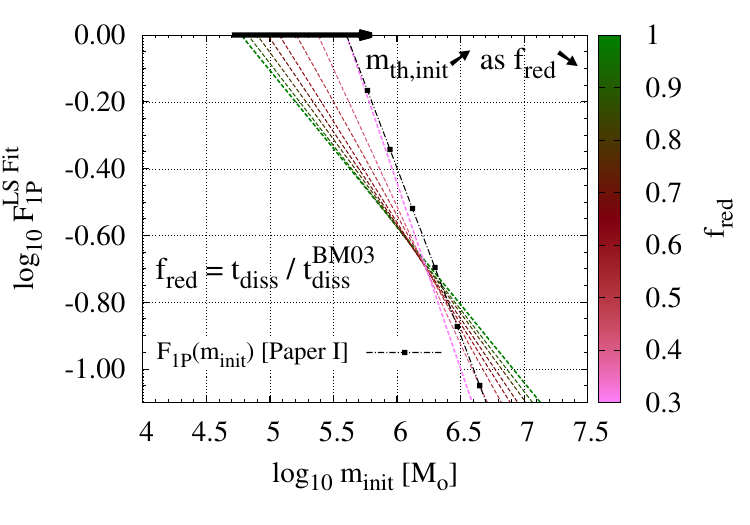}  
\caption{Least-square fits to the 12 disk \gcs for different reductions $f_{red}$ of the \citet{bm03} cluster dissolution time-scale.  The smaller $f_{red}$, the shorter the cluster dissolution time-scale $t_{diss} = f_{red} t_{diss}^{BM03}$, the steeper the slope of the least-square fit and the higher the initial mass threshold $m_{th,init}$ that clusters must have to host  polluted stars at secular-evolution onset (values of $m_{th,init}$ are given by the intersections betwen fits and top $x$-axis).  The greenest ($f_{red}=1.0$) and pinkest ($f_{red}=0.3$) lines are the solid orange lines in the middle and bottom panels of Fig.~\ref{fig:disk}, respectively.  
}
\label{fig:degen} 
\end{figure} 

In all three panels, the 12 disk clusters are well correlated.  Their correlation coefficient, $r_{Disk}$, and their least-squares fit (solid dark-orange line) are given in each panel.  As explained in Sec.~\ref{sec:minit}, the fit slope steepens and its intercept increases from the top to the bottom panels, as cluster mass losses are corrected for.  The fit obeys $\log_{10}(F_{1P})=(-0.320\pm0.017)\log_{10}(m_{prst})+(1.281\pm0.091)$ in the present-day space, $\log_{10}(F_{1P})=(-0.467\pm0.026)\log_{10}(m_{init})+(2.229\pm0.148)$ when $f_{red}=1.0$ and $\log_{10}(F_{1P})=(-1.110\pm0.091)\log_{10}(m_{init})+(6.217\pm0.550)$ when $f_{red}=0.3$.  In the latter case (bottom panel of Fig.~\ref{fig:disk}), the reduction factor is that assumed in Paper~I and the least-squares fit almost coincides with the relation $F_{1P}(m_{init})$ from Paper~I.       

The tight correlation shown by the disk clusters in the $(m_{init},F_{1P})$ space allows us to track and visualize the degeneracy between, on the one hand, the initial mass threshold $m_{th,init}$ that clusters must have to host multiple populations at secular-evolution onset and, on the other hand, the reduction factor $f_{red}$ applied to the \citet{bm03} cluster dissolution time-scale $t_{diss}^{BM03}$.   
The stronger the reduction of $t_{diss}^{BM03}$ (i.e., the smaller $f_{red}$), the higher the inferred cluster initial masses and the higher the threshold $m_{th,init}$.  This is shown in Fig.~\ref{fig:degen}, where each line depicts the least-square fit to the disk clusters for a given color-coded reduction factor $f_{red}$ (see palette).  
We stress again that this threshold stands at secular-evolution onset.  The threshold required {\it to form} polluted stars is higher by a factor $1/F_{bound}^{VR}$.  We shall discuss the threshold further in Sec.~\ref{sec:ninst}

\subsection{Accreted clusters}
\label{ssec:accr}
\begin{figure}
\includegraphics[width=0.49\textwidth, trim = 0  20 0 0]{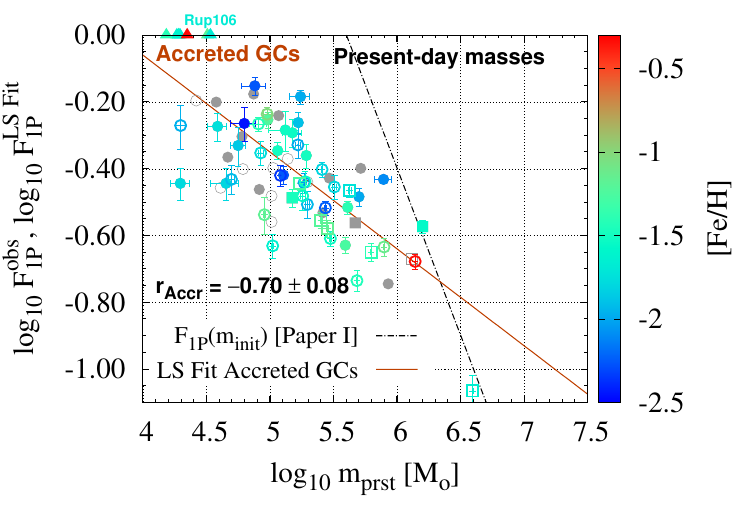}  \\
\includegraphics[width=0.49\textwidth, trim = 0  60 0 0]{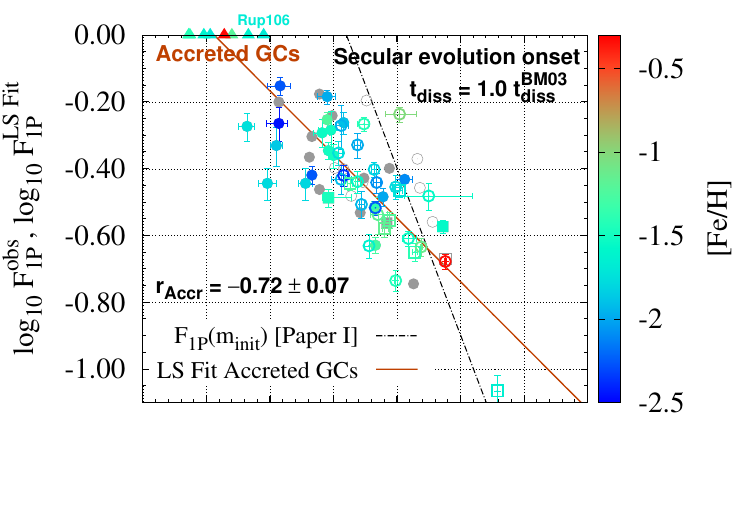}  \\
\includegraphics[width=0.49\textwidth, trim = 0   0 0 0]{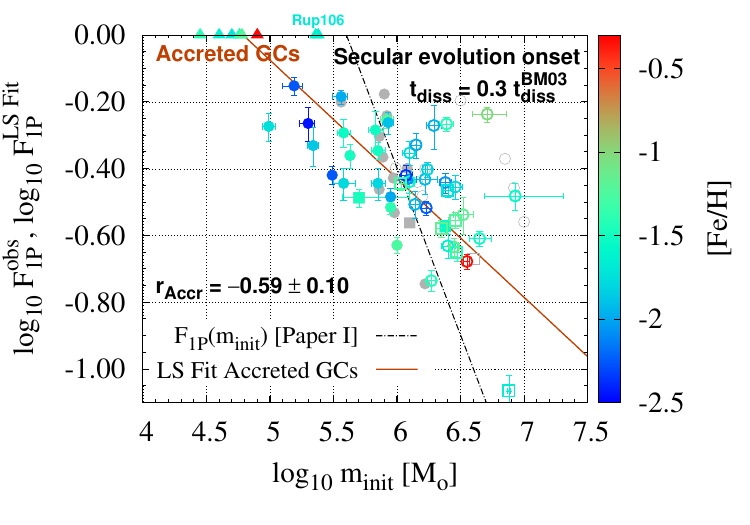}
\caption{
Top panel: distribution of our globular cluster sample in the $(m_{prst},F_{1P})$ space with the 46 accreted clusters of \citet{mass19} color-coded according to their  mean metallicity (see palette).  Each panel shows the accreted-group correlation coeffficient $r_{Accr}$ and least-square fit (solid dark-orange line; the fit covers the Type~I and Type~II clusters, but exclude the single-population clusters).  The black dashed-dotted line is the initial relation from Paper~I, $F_{1P}(m_{init})=4\cdot10^5\Ms m_{init}^{-1}$.  Middle panel: same for the $(m_{init},F_{1P})$ space with  $f_{red}=1.0$.  Bottom panel: same for the $(m_{init},F_{1P})$ space with $f_{red}=0.3$. 
}
\label{fig:accr} 
\end{figure} 
Figure~\ref{fig:accr} is the counterpart of Fig.~\ref{fig:disk} for all accreted clusters, now those  color-coded for mean metallicity.  
The dwarf galaxies identified so far for having contributed \gcs to the Galaxy are the progenitor of the Helmi stream \citep{hel99}, {\it Gaia}-Enceladus \citep{belokurov18,helm18}, Sequoia \citep{mye19} and the currently merging Sagittarius dwarf galaxy \citep{iba94}.  

\citet{mass19} identified two additional groups of accreted clusters that remain unassociated to well established merger events: firstly, a group of clusters that they tentatively assign to a low-energy structure with near-zero $L_Z$ angular momentum (pink symbols in their Fig.~2; labels 'L-E' in their Table~A.1); secondly, a group of high-energy (i.e., loosely bound to the Galaxy) clusters whose scattered distribution in the space of integrals-of-motion discards a common origin (cyan symbols in their Fig.~2; labels 'H-E' in their Table~A.1).  

The eight single-population clusters are all accreted and six of them are high-energy clusters \citep[][their Sec.~6; see also our Fig.~\ref{fig:origin}]{mil20}.  In the present-day space, they are offset from the least-square fit to the accreted multiple-population clusters (top panel of Fig.~\ref{fig:accr}).  In the $(m_{init},F_{1P})$ space, however, most of them extend the fit, despite the fit not taking them into account (middle and bottom panels).  When $f_{red}=0.3$ (bottom panel), two exceptions are Rup~106 and Pal~14, the most massive single-population clusters, which depart significantly from their single-population siblings.  Rup~106 is intriguing because it behaves like a multiple-population cluster in the metallicity-vs-compactness space \citep[Fig.~4 in][]{hua24}.  \citet{par24b} therefore asks whether Rup~106 could have formed as a multiple-population cluster, with a centrally concentrated pristine population and an outer polluted population.  Following the tidal stripping of its outer 2P stars, Rup~106 would have been reduced to a cluster of pristine stars only (its pericentric distance $D_{peri}=4.6$\,kpc is much shorter than its current Galactocentric distance $D_{gal}=18.5$\,kpc).  If Rup~106 initially follows the least-squares fit of the multiple-population clusters, the bottom panel of Fig.~\ref{fig:accr} suggests an initial pristine-star fraction of $F_{1P} \simeq 0.6$.  

We now consider the low-energy and Gaia-Enceladus groups separately.  
   
\subsubsection{Massari et al. 's Low-Energy group}
\label{sssec:LE}
        
\begin{figure*}
\includegraphics[width=0.49\textwidth, trim = 0  20 0 0]{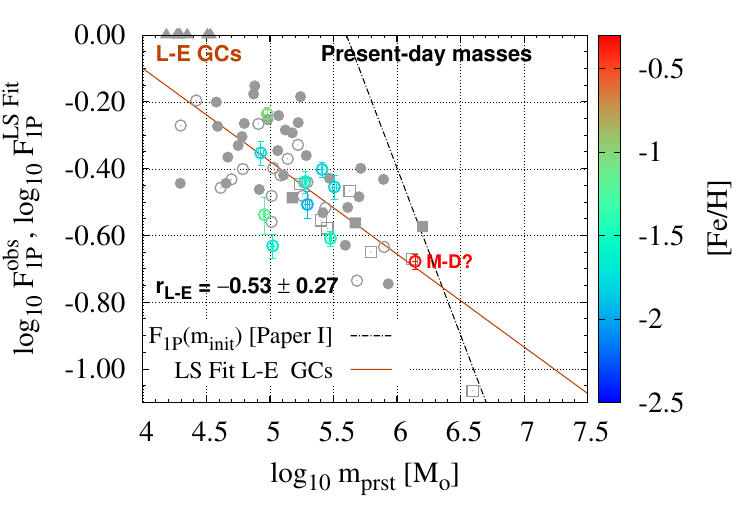} 
\includegraphics[width=0.49\textwidth, trim = 0  20 0 0]{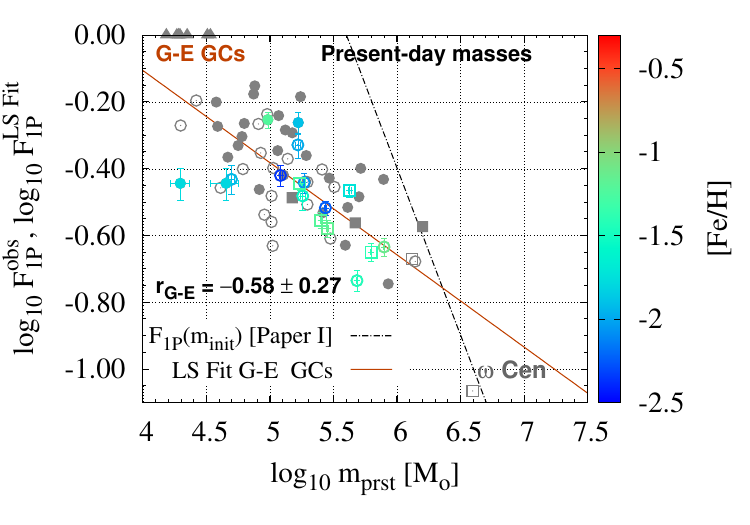}  \\
\includegraphics[width=0.49\textwidth, trim = 0  60 0 0]{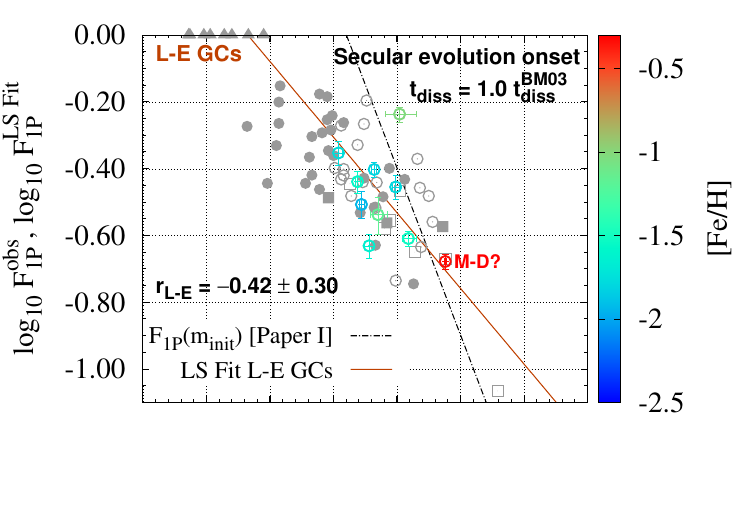}
\includegraphics[width=0.49\textwidth, trim = 0  60 0 0]{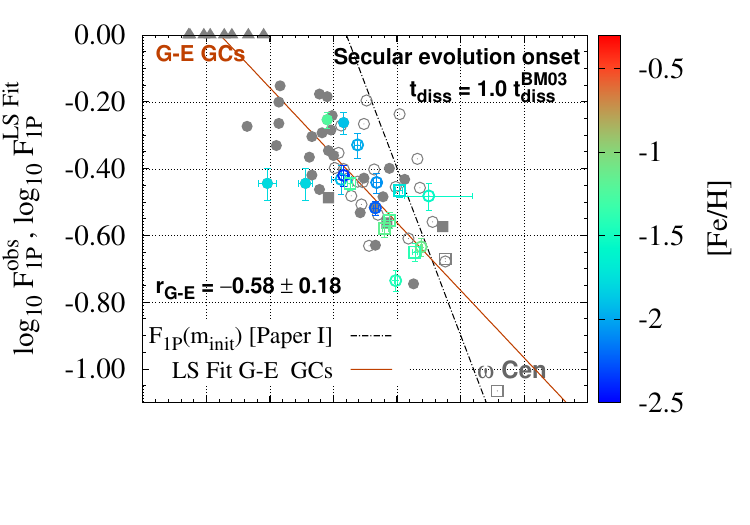}  \\
\includegraphics[width=0.49\textwidth, trim = 0  00 0 0]{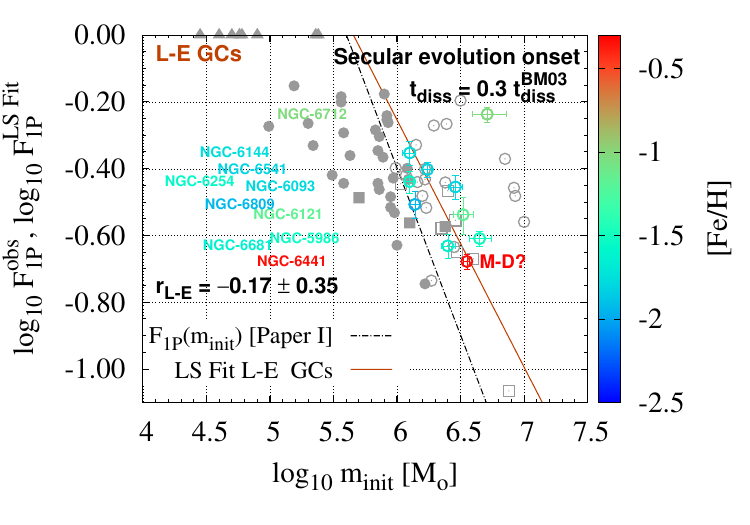}
\includegraphics[width=0.49\textwidth, trim = 0  00 0 0]{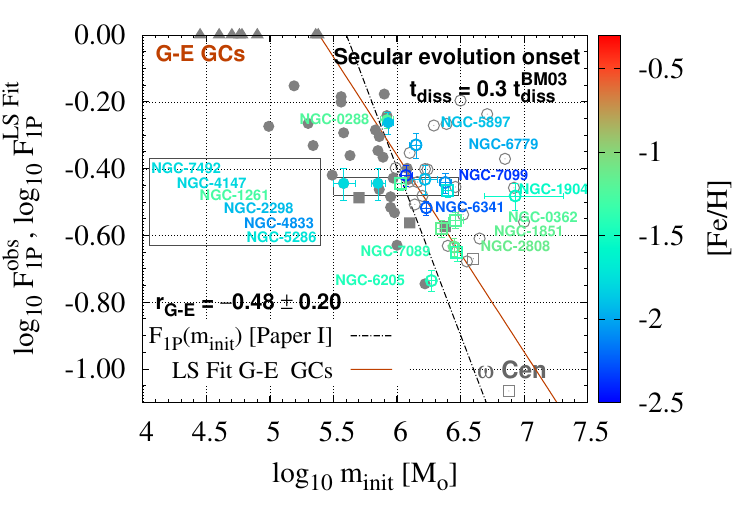}
\caption{Same as Fig.~\ref{fig:disk} for the low-energy clusters (left) and Gaia-Enceladus clusters (right).  In the bottom-right panel, the 6 cluster symbols with similar $F_{1P}$ values in the grey rectangle have their IDs given in the frame to its left.}
\label{fig:LEGE} 
\end{figure*} 

The low-energy group of \citet{mass19}, highlighted in the left panels of Fig.~\ref{fig:LEGE}, consists of Type~I inner clusters only (i.e., all colored symbols in Fig.~\ref{fig:LEGE} left panels are open circles).  \citet{mass19} show this group to be highly clustered in the space of integrals-of-motion (pink points in their Fig.~2), with a tight age-metallicity relation (bottom right panel in their Fig.~4).   They suggest the low-energy group to stem from the accretion of a large hitherto unknown galaxy, at least as massive as Gaia-Enceladus \citep[see also][]{garc23}. 

The limited region occupied by the low-energy clusters in the $(m_{init},F_{1P})$ space is consistent with their arising from a single progenitor, although NGC~6712 ($\log_{10}(F_{1P})\simeq-0.25$) looks like an outlier.  At the bottom of the distribution, NGC~6441 is, according to \citet{mas23}, an in-situ cluster rather than an accreted cluster.  Given its apocentric distance $D_{apo} = 4.7>3.5$\,kpc, we have labelled it 'M-D?' in the bottom left panel of Fig.~\ref{fig:LEGE}.  

\subsubsection{The Gaia-Enceladus group}
\label{sssec:GE}
The Gaia-Enceladus group of \citet{mass19}, highlighted in the right panels of Fig.~\ref{fig:LEGE}, consists of 12 Type~I and 5 Type~II clusters.  \citet{mil20} already noticed that this merger-event progenitor was especially efficient at producing Type~II clusters, about half of them being associated to Gaia-Enceladus.  

The next section quantifies and compares the dispersion of the disk, low-energy and Gaia-Enceladus  groups in the $(m_{prst},F_{1P})$ and $(m_{init},F_{1P})$ spaces.

\section{The dispersion in $(m_{init},F_{1P})$ space to probe cluster formation conditions}
\label{sec:hdisp}

Since clusters are assumed to evolve at constant $F_{1P}$, the dispersion of a given cluster group around its least-squares fit in the $(m_{init},F_{1P})$ space is assigned to cluster initial-mass variations only.  In turn, initial-mass variations reflect bound-fraction variations (recall that $m_{init}=F_{bound}^{VR}m_{ecl}$), themselves driven by variations in cluster-formation conditions, e.g.~in the \sfe achieved by cluster-forming clumps.  

\begin{figure}
\includegraphics[width=0.49\textwidth, trim = 0  0 0 0]{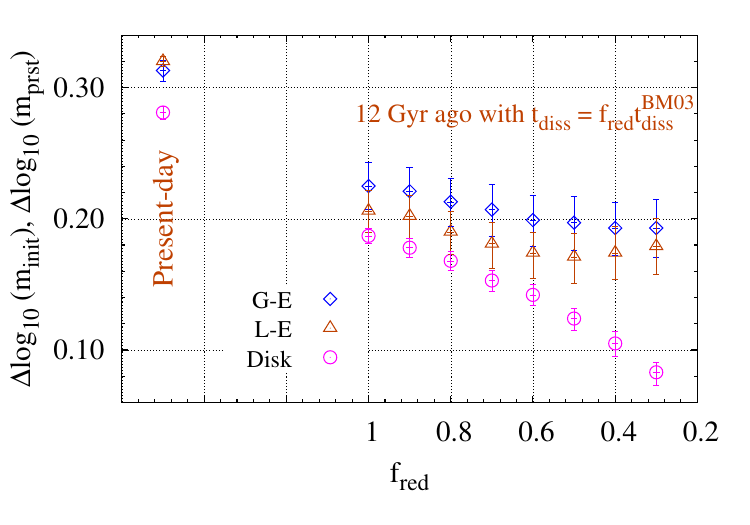}
\caption{Dispersion $\Delta\log_{10}(m_{init})$ of the data points in the $(m_{init},F_{1P})$ space around their respective least-squares fit, in dependence of the reduction factor $f_{red}$ applied to the \citet{bm03} cluster dissolution time-scale.  Results are shown for the disk, low-energy and Gaia-Enceladus cluster groups.  The three points to the top left show the dispersion $\Delta\log_{10}(m_{prst})$ in the present-day space}    
\label{fig:dhoriz} 
\end{figure} 

To quantify the dispersion in logarithmic initial mass of a cluster group, we calculate the mean of the absolute horizontal distance between the data points and their least-squares fit, $\Delta \log_{10}(m_{init})$.  We do so for each value of $f_{red}$ for which a set of cluster initial masses has been obtained.  

Results for the disk, low-energy and Gaia-Enceladus groups are shown in Fig.~\ref{fig:dhoriz}, where moving rightward shortens the cluster dissolution time-scale.  For the sake of comparison, results for the present-day distribution $(m_{prst},F_{1P})$ are depicted in the top left corner of the plot.  
The distributions of data points around their respective least-squares fit are tighter in the initial space than in the present-day space, regardless of the adopted reduction factor $f_{red}$.  In other words, secular evolution disperses the data points, as expected (see also Fig.~\ref{fig:loglog}).  

In the $(m_{init}, F_{1P})$ space, the disk clusters constitute the tightest group, especially at short cluster dissolution time-scale (see also Fig.~\ref{fig:disk}).  By  contrast, the accreted groups are more loosely distributed and their dispersion $\Delta \log_{10}(m_{init})$ does not depend sensitively on $f_{red}$.  These properties can be explained in terms of \stf efficiency.  

\begin{figure}
\includegraphics[width=0.49\textwidth, trim = 0 65 0 0]{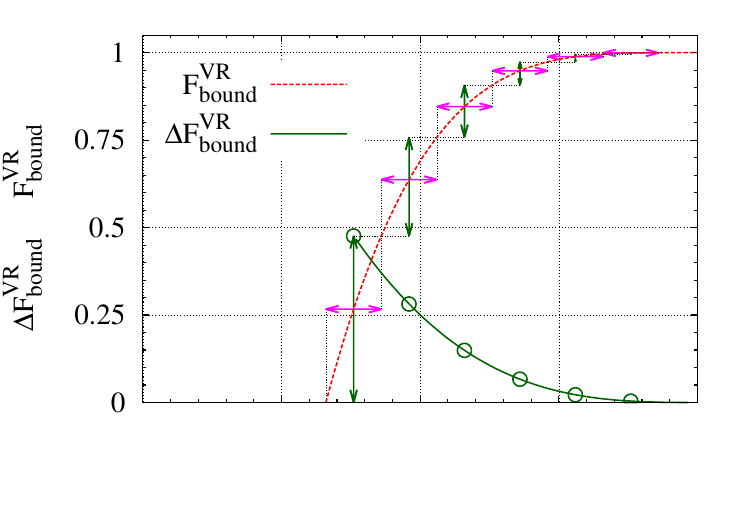}  \\
\includegraphics[width=0.49\textwidth, trim = 0  0 0 0]{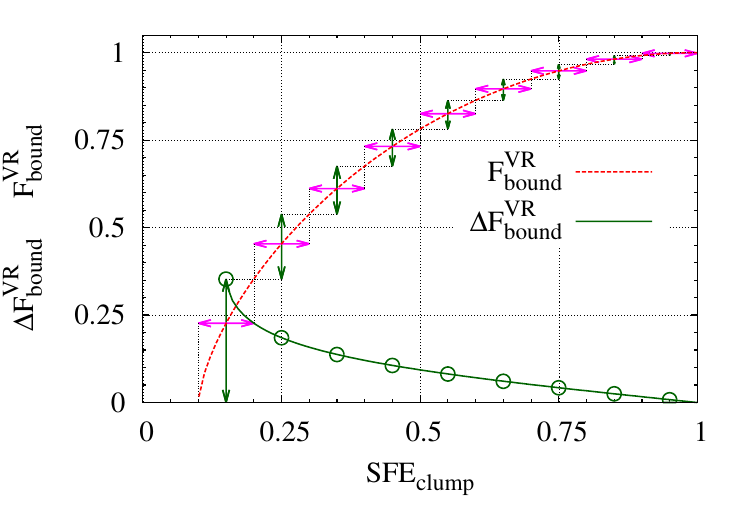}
\caption{Top panel: Cluster bound fraction $F_{bound}^{VR}$ after residual \sfing gas expulsion in dependence of cluster-progenitor \sfe $SFE_{clump}$ \citep[dashed red line from Fig.~1 in][]{par07}.  For a fixed range $\Delta SFE_{clump} = 0.1$ (horizontal arrows), the vertical arrows and the solid green line depict the corresponding bound-fraction range $\Delta F_{bound}^{VR}$.  The density profiles of the stars and gas in embedded clusters are assumed to have the same shape.  Bottom panel: same as top panel for clusters forming spatially concentrated in their embedding gas \citep[dashed red line from Fig.~3 in][]{ada00}.  }    
\label{fig:sfefb} 
\end{figure} 

The dashed red lines in both panels of Fig.~\ref{fig:sfefb} show the stellar mass fraction $F_{bound}^{VR}$ that clusters retain by the end of violent relaxation, as a function of their  progenitor \stf efficiency, $SFE_{clump}$.  Two cases are considered: 
the \sfe is either (1) constant all through a cluster-forming clump \citep[top panel; from Fig.~1 in][]{par07} or (2)~higher in clump inner regions than in clump outskirts \citep[bottom panel; the dashed red line is the dotted line in Fig.~3 of][]{ada00}, which helps clusters survive gas expulsion better than in the first case (see figs~1 and 10 in \citealt{par13} and \citealt{ada00}).  
The double-headed horizontal magenta arrows depict a fixed range $\Delta SFE_{clump}=0.1$ in \stf efficiency, and the vertical green arrows show the corresponding bound-fraction ranges, $\Delta  F_{bound}^{VR}$.  $\Delta F_{bound}^{VR}$ is also shown in dependence of $SFE_{clump}$ (solid green lines with open circles).  

When the \sfe $SFE_{clump}$ inside cluster-forming clumps is (globally) high, not only is the bound fraction $F_{bound}^{VR}$ high (dashed red line), it is also weakly scattered (solid green line).  As a result, the relation $F_{1P}(m_{init}) $ between pristine-star fraction and cluster initial mass does not differ much from its embedded counterpart $F_{1P}(m_{ecl})$ (since  $m_{init} = F_{bound}^{VR} ~ m_{ecl} \lesssim m_{ecl}$).  
Toward lower \stf efficiencies, however, the range of bound fractions $\Delta  F_{bound}^{VR}$ shoots up.  This holds regardless of the gas-versus-stars spatial distributions. 
As a result, even a limited range in \stf efficiencies $\Delta SFE_{clump}$ turns a one-to-one embedded-cluster track $F_{1P}(m_{ecl})$ into a dispersed initial distribution $F_{1P}(m_{init})$.        

Building on Fig.~\ref{fig:sfefb}, Fig.~\ref{fig:dhoriz} can be interpreted as protoglobular-cluster clumps of the young Galactic disk achieving significantly higher \stf efficiencies than their siblings in dwarf galaxies.  
If the \sfe in disk clumps is as high as, say 85\,\%, then variations in other clump properties (e.g.~gas expulsion time-scale, virial ratio, external tidal field, ...) hardly matter since their poor gas content implies the robustness of their clusters following gas expulsion.  By contrast, clumps in dwarf galaxies would achieve lower \stf efficiencies, resulting in wider bound-fraction ranges and more dispersed distributions in the $(m_{init},F_{1P})$ space.      

High-resolution simulations of massive-cluster formation actually show that cluster-forming clumps with high surface density achieve higher \stf efficiencies than their low surface-density counterparts \citep[e.g.,][and references therein]{pol23}.  Star formation efficiencies as high as 85\,\% can be achieved, an effect assigned by \citet{pol23} to the short time-scale for gravitational collapse and to the clump deep potential.  As a result, stellar feedback starts too late and is too weak to expel the residual \sfing gas.

The difference between the $\Delta log_{10}(m_{init})$-dispersions of the disk and accreted groups may also reveal more uniform cluster-formation conditions in the young Galactic disk than in accreted systems (e.g.~similar \stf efficiencies but a wider dispersion around the mean $SFE_{clump}$ in  dwarf galaxies than in the Galactic disk).

An alternative explanation to Fig.~\ref{fig:dhoriz}, however, is that the mass of the most massive and inner clusters is overestimated as Eq.~\ref{eq:minit} does not correct for the impact of dynamical friction \citep{bt87}.  When dynamical friction and the associated decay of the cluster orbit are neglected, cluster initial masses are overestimated since based on orbits that are closer to the Galactic center today than in the past.  Therefore, when a cluster group is dominated by inner clusters, its $\Delta log_{10}(m_{init})$-dispersion could be artificially inflated as a result of overestimating the initial mass of the clusters closest to the Galactic center.  This may be the reason why $\Delta\log_{10}(m_{init})$ for the low-energy and Gaia-Enceladus groups is "plateau-ing" as $f_{red}$ decreases, that is, as initial mass estimates get higher and, therefore, dynamical friction stronger.  Because of their larger Galactocentric distance, disk clusters are immune to this effect.  Refined modeling of cluster mass losses is therefore required to assess the actual  difference in $\Delta log_{10}(m_{init})$-dispersion between the disk and accreted groups.

\section{Cluster metallicity and pristine-star fraction}
\label{sec:FeH}
Virtually all Galactic \gcs present the O-Na anticorrelation. 
The Mg-Al anticorrelation, however, is rarer \citep[compare Figs~4 and 6 in][]{car09b}, showing up in high-mass and/or low-metallicity clusters \citep[see Fig.~5 in][]{panci17}.  Specifically, the observed amount of Al produced in multiple-population clusters and their depletion in O are both correlated with a linear combination of cluster metallicity and absolute visual magnitude (see Fig.~20 in \citealt{car09a} and Fig.~13 in \citealt{car09b}, respectively), with the stronger O-depletions and Al-enhancements in brighter and more metal-poor globular clusters.  \citet{schia24} find a clear anticorrelation between the mean metallicity of Galactic \gcs and the extension of their Mg-Al anticorrelation (their Fig.~11).  Metal-poor clusters are also more prone on hosting multiple populations than metal-rich ones (Fig.~4 in \citealt{hua24}, Fig.~2 in \citealt{charb16}).  

A metallicity dependence of the multiple-population phenomenon may be expected from metal-poor protoclusters either forming more polluting elements or retaining them better.  
Clusters can indeed retain better the ejecta of metal-poor massive stars because of their weaker stellar winds \citep{kud87,vink11}, while, in the AGB scenario, more metal-poor AGB stars produce more Al at the expense of Mg \citep{ven16}.  

It is thus tempting to enquire whether, in Figs~\ref{fig:disk} and \ref{fig:LEGE}, more metal-poor clusters {\it of a given initial-mass range} host a greater fraction of polluted stars  (i.e., reach lower $F_{1P}$ values).  Our aim is {\it not} to search for a $F_{1P}$-[Fe/H] correlation here.  None was found by \citet{mil17}. The question is: for any given cluster initial-mass range, are the data points above the least-squares fits rather reddish, and those below the least-squares fits rather bluish?  This effect would be akin to the metallicity-dependent extent of the Mg-Al anticorrelation shown by the right panels of Fig.~5 in \citet{panci17}.  

At first glance, none of the disk (Fig.~\ref{fig:disk}), Gaia-Enceladus or low-energy (Fig.~\ref{fig:LEGE}) group reveals such a trend.  
But should it be the case for the $F_{1P}(m_{init})$ relation?  Let us assume that the metallicity of the intra-cluster gas does influence the pristine-star fraction.  This impacts the embeded-cluster relation $F_{1P}(m_{ecl})$.  Can such a metallicity imprint survive violent relaxation?  To illustrate that this may not be the case, let us consider the model of \citet{dib11} in which more metal-poor clusters experience longer \stf time-spans as a result of their OB-stars blowing weaker winds (their Figs~12-13).  We stress that \citet{dib11} 's  model was not developed specifically for \gc formation: their model cluster masses are lower and their metallicities higher than for most forming globular clusters (their most massive clump is $2 \cdot 10^5\,\Ms$ and their lowest metallicity is a tenth of the solar metallicity).  Our applying of their conclusions is therefore purely qualitative.  Because 2P stars form toward the end of cluster formation rather than early on, the longer \stf time-span of more metal-poor clusters should lead to higher 2P-star fractions, as might be intuitively expected (but see below).  This is sketched in Fig.~\ref{fig:fehsketch} where plain blue circles represent newly-formed metal-poor clusters with polluted-star fractions higher (i.e., $F_{1P}$ is lower) than the averaged $F_{1P}^{Aver}(m_{ecl})$ of the group they belong to.  In contrast, the pristine-star fraction of their metal-rich siblings (plain red squares) is higher, as a consequence of their shorter \stf time-span.   

\begin{figure}
  
\includegraphics[width=0.49\textwidth, trim = 0  0 0 0]{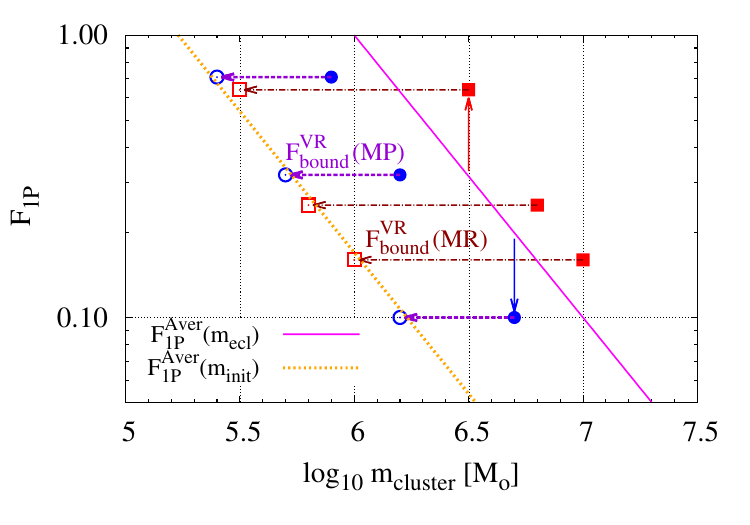}
\caption{Qualitative depiction of how a metallicity-dependent embedded-cluster relation $F_{1P}(m_{ecl})$ (solid magenta line and plain symbols) can evolve into a metallicity-independent initial relation $F_{1P}(m_{init})$ (dotted orange line and open symbols).  In the case depicted here, gas-embedded metal-poor clusters (blue plain circles) have lower pristine-star fractions than an average $F_{1P}^{Aver}(m_{ecl})$ relation, while the pristine-star fraction of their metal-rich counterparts (red plain  squares) is higher.  But with metal-poor clusters retaining a greater star fraction $F_{bound}^{VR}$ than metal-rich clusters (horizontal arrows: dashed-dotted and dashed for the metal-rich ('MR') and metal-poor ('MP') cases, respectively), the initial relation $F_{1P}(m_{init})$ that emerges after violent relaxation is left unimpacted by metallicity.
}    
\label{fig:fehsketch} 
\end{figure} 

Because of their longer \stf time-span, metal-poor clusters also reach higher \stf efficiencies \citep[Fig.~14 in][]{dib11} and retain therefore a greater fraction $F_{bound}^{VR}$ of their stars after gas expulsion (see our Fig.~\ref{fig:sfefb}).  Figure~\ref{fig:fehsketch} shows how both effects could then cancel each other, as the lower bound fractions of metal-rich clusters compensate for their higher pristine-star fractions at formation, leading to an initial relation $F_{1P}(m_{init})$ void of any metallity influence.  Depending on the variations of $F_{1P}(m_{ecl})$ and $F_{bound}^{VR}$ with the \sfing gas metallicity, depending on the slope of $F_{1P}(m_{ecl})$, one could imagine scenarios in which the imprint of metallicity on $F_{1P}(m_{ecl})$ is, following violent relaxation, weakened, cancelled (as in Fig.~\ref{fig:fehsketch}) or, even, reversed.
That metallicity does not seem to impact the $F_{1P}(m_{init})$ relations obtained in our Figs.~\ref{fig:disk}-\ref{fig:LEGE} may therefore not be so surprising.  Yet, it does not imply that metallicity plays no role in the formation of cluster multiple populations.  Its influence could actually be trickier than intuitively thought.

Consider for instance the recent model of \citet{gieles25}, in which accreting Extremely Massive Stars (aEMSs, mass $\sim 10^3-10^4\,\Ms$) are the dominant polluters of forming globular clusters.  In their model, more massive and more metal-poor clusters present stronger Mg depletions and Al enhancements due to the higher aEMS central temperatures \citep[Fig.~18 in][]{gieles25}.  This agrees with Carretta 's empirical bivariate relations quoted earlier in this section.  Yet, their predicted polluted-star fraction $1-F_{1P}$ slightly {\it increases} with metallicity (their Fig.~15), in contrast with the intuitive expectation that metal-poor clusters would form a greater fraction of polluted stars.  This effect stems from their aEMSs reaching their asymptotic mass (when accretion and wind mass-loss rates balance each other) on a shorter time-scale $\tau_{\infty}$ in metal-richer environments ($\tau_{\infty} \propto Z^{-0.3}$, see end of their Sec.~5.3)\footnote{The reader may be puzzled that the models of \citet{dib11} and \citet{gieles25} lead to conflicting interpretations.  Both cluster-formation models differ on more than just their forming-cluster mass range. \citet{dib11} consider fully preassembled protostellar cores and protocluster clumps (no inflows, no accreting forming stars) and their stellar mass spectrum covers up to the O-star regime.  In contrast, \citet{gieles25} consider forming clusters fed by converging flows driven by supersonic turbulence, which allows their stars to become more massive than their prestellar cores, eventually yielding the aEMSs}.  \\

In the $(m_{init}, F_{2P}=1-F_{1P})$ data space, however, they find no significant correlation (one should keep in mind here that \citet{gieles25} 's model pertains to the cluster-formation stage, while their data points refer to the initial cluster masses.  It remains to be seen how violent relaxation could impact their model).  For instance, the bottom-left panel of their Fig.~15 shows that neither the metal-rich Galactic \gcs nor the metal-poor ones tend to cluster toward the top of the plot (see their symbols color-coded for metallicity).  That is, in a given initial mass range, neither the metal-rich clusters nor the metal-poor ones tend to host a greater fraction of 2P stars.  This agrees with our Figs~\ref{fig:disk} and \ref{fig:LEGE} where the reddish symbols (most metal-rich clusters) show no tendency of clustering either below or above the least-squares fits.  \\

To summarize: while the impact of metallicity on the Mg-Al anticorrelation is well established, it is much less straightforward to predict how it affects the initial relation $F_{1P}(m_{init})$.  Should $F_{1P}(m_{init})$ be metallicity-dependent, it is also not so straightforward to unearth this effect among currently available data sets, for the following reasons:  \\  

(i)~When the 1P and 2P populations are not well mixed, measured $F_{1P}$ are sensitive to the selected cluster region.  Consider NGC~104 (47~Tuc).  \citet{mil17} inferred $F_{1P} = 0.18$ based on HST observations of its inner regions.  Yet, ground-based observations reaching out to the outer regions, but ignoring its core, have doubled this value \citep[$F_{1P}=0.40$,][top-row panel of their Fig.~3]{jang25}.  The cluster average thus lies in-between. It is probably no coincidence that \citet{car10a} find, based on the O-Na anticorrelation, a pristine-star fraction $F_{1P}=0.27$ for 47Tuc (their Table~2).  Pristine-star fractions obtained from combining HST observations of the crowded central regions and ground-based observations maximizing the cluster spatial coverage are urgently needed \citep[e.g.][]{leit23} (see also Sec.~\ref{sec:phsp}).\\

(ii)~The small number of clusters in each group hinders the search of an [Fe/H] impact on its  $F_{1P}(m_{init})$ relation.  It should be kept in mind that our cluster census for the disk, low-energy and Gaia-Enceladus groups remain incomplete as we miss the clusters with no $F_{1P}$-estimate.  Additionnally, the assignment of clusters to a given group is sometimes ambiguous.  For instance, in the bottom panel of Fig.~\ref{fig:disk}, the four most metal-poor disk clusters of \citet{mass19} (bluest symbols) are assigned to either Gaia-Enceladus or the low-energy group by \citet[][their Table~A1]{callingham22}.  Conversely, NGC6441 (reddest symbol in the bottom-left panel of Fig.~\ref{fig:LEGE}) is considered a disk cluster by \citet{callingham22} and \citet{mas23}, but a low-energy one by \citet{mass19}.  Incomplete and/or "contaminated" cluster sample do not facilitate the search for a metallicity-dependent $F_{1P}(m_{init})$ relation, especially when the ambiguous clusters are those with high or low metallicities.  \\

(iii)~Our initial mass estimates are more uncertain than suggested by their error bars.  The latter do not account for the time-varying gravitational potential in which clusters evolve.  The Milky-Way potential has changed with time and, when it comes to accreted clusters, these swapped the potential of their natal dwarf galaxy for that of the Milky Way.  This latter effect, however, should not be significant for the Gaia-Enceladus group.  With a merger dating back to 10\,Gyr ago \citep{helm18, gall19}, its \gcs have evolved mostly in the Milky Way potential.  As for the low-energy clusters, were they accreted from a Kraken-like galaxy as suggested by \citet{mass19}, their accretion would date back to an even earlier time \citep{garc23}\footnote{Note that the low-energy progenitor has also been coined {\it Koala} by \citet{forbes20}}.\\

(iv)~To ignore the impact of dynamical friction on the most inner massive clusters equates with overestimating their initial mass (see Sec.~\ref{sec:hdisp}).  \\

(v)~Finally, pristine-star fractions vary following the preferential removal of either 1P or 2P stars, an effect our model must still include (Parmentier, {\it in prep.}).  \\ 

Refined modeling and more accurate data are therefore required prior to concluding definitively.  Given the uncertainties affecting $F_{1P}$ and $m_{init}$, the search of metallicity as a second driver, after the cluster mass, of the pristine-star fraction $F_{1P}$ is not yet over and requires an open-mind.  While more metal-poor stellar polluters produce more Al through hotter H-burning, metal-rich stellar polluters cast more of their processed mass in the intra-cluster gas owing to their stronger winds \citep[see e.g. Sec.~5.4 in][]{gieles25}.

\section{$F_{1P}$ Uncertainties (again): Photometry versus Spectroscopy}
\label{sec:phsp}

The pristine-star fractions of our cluster sample \citep[e.g.][]{mil17} stem from stellar fluxes through ultraviolet passbands that contain the NH and CN molecular bands \citep[Fig.~1 in][]{pio15}.  That is, the multiple-population classification that we use builds on the nitrogen abundance \citep[Fig.~27 in][]{marino19}.  Recently, however, \citet{carrbrag24} have shown that the N-based classification, be it from HST photometry or low-resolution spectroscopy, does not always agree with the sodium abundances from high-resolution spectroscopy (their Figs.~8 and 11).  For instance, some stars labelled 1P based on their location in the chromosome map show the large Na abundance expected for polluted stars, and vice-versa.  A similar effect was identified by \citet{smith13} in NGC5904, whose CN-strong and CN-weak giants overlap in their sodium abundances (their Fig.~8).  This missclassification, which, depending on the host cluster, affects from 4 to 33\,\% of the stars \citep[Table~4 in][]{carrbrag24}, may result from the N and Na enhancements reflecting different phases of the \gc complex pollution history \citep{carrbrag24}.

In most cases, N-based indices tend to overestimate the pristine-star fraction, especially in the low cluster-mass regime, be it the present-day or the initial mass \citep[Fig.~12 and Sec.~4 of][]{carrbrag24}.  The pristine-star fractions in the top part of the $F_{1P}(m_{prst})$ and $F_{1P}(m_{init})$ relations may thus be too high, which would imply shallower $F_{1P}(m_{init})$ relations and lower cluster-mass thresholds than suggested by our Figs~\ref{fig:disk}-\ref{fig:LEGE}.     

Additionally, toward lower metallicity, UV-based pseudo-colours increasingly fail to distinguish between polluted and pristine stars \cite[Fig.~15 in][]{lee24}.  This is likely a consequence of the weakening spectral signature of the bi-metal CN molecule at low metallicity.  At low [Fe/H], \citet[][]{carrbrag24} therefore find that N-based pristine-star fractions also tend to be higher than those derived from high-resolution Na abundances (their Figs~7 and 10).  Because our Figs.~\ref{fig:disk}-\ref{fig:LEGE} do not reveal any trend with cluster metallicity (Sec.~\ref{sec:FeH}), we do not expect any systematic effect on our results.

\section{Generalizing the cluster-mass threshold for 2P-star formation}
\label{sec:ninst}
The $F_{1P}(m_{ecl})$ relation of Paper~I describes clusters that self-pollute completely and instantaneously once their developing stellar mass reaches a fixed threshold $m_{th}$.  It obeys $F_{1P}=(m_{th}/m_{ecl})^\psi$ with $\psi=1$, hence $F_{1P} = (m_{th,init}/m_{init})$ for a fixed bound fraction $F_{bound}^{VR}$.  By contrast, the relations of Fig.~\ref{fig:degen} are power laws with $\psi \neq 1$.

\begin{figure}
\includegraphics[width=0.49\textwidth, trim = 0  20 0 0]{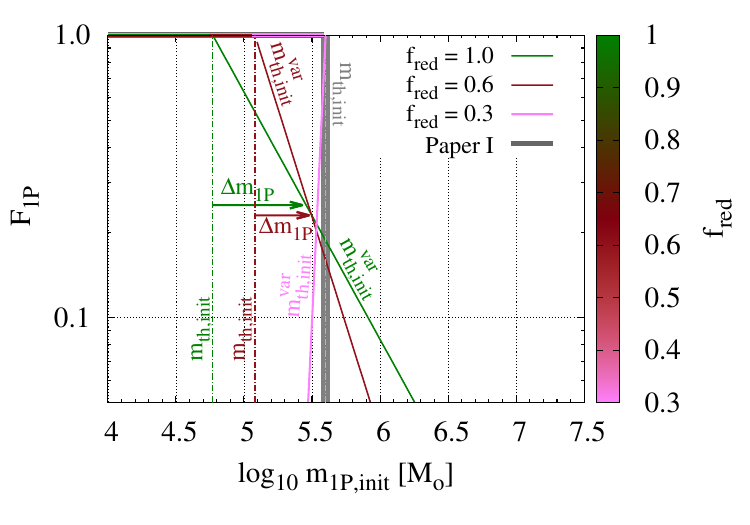}
\caption{
Relation between the pristine-star fraction $F_{1P}$ of clusters and their mass in 1P stars $m_{1P,init}$ at secular-evolution onset.  They follow from the relations between pristine-star fraction and initial mass for the disk clusters (Fig.~\ref{fig:degen}, same color coding).  Assuming a constant $F_{bound}^{VR}$, the solid lines for $f_{red}=1.0$ and $f_{red}=0.6$ illustrate  either a mass-dependent threshold $m_{th,init}^{var}/F_{bound}^{VR}$ for 2P-star formation combined with instantaneous cluster pollution, or the combination of a fixed mass threshold $m_{th,init}/F_{bound}^{VR}$ (dashed-dotted vertical lines) with prolonged 1P-star formation (horizontal arrows labelled $\Delta m_{1P}$). 
When $f_{red}=0.3$, more massive clusters experience a lower mass theshold for 2P-star formation.  The thick vertical gray line is Paper~I model for instantaneous cluster pollution (see also top panels of Figs~4-5 in Paper~II).   
}
\label{fig:ninst} 
\end{figure}

To grasp what a logarithmic slope $-\psi \neq -1$ implies, let us consider Fig.~\ref{fig:ninst}, which shows the pristine-star fraction $F_{1P}$ in dependence of $m_{1P,init}$, the mass in pristine stars in clusters at secular-evolution onset.  The thick vertical solid line is the fixed threshold $m_{th,init}=4\cdot 10^5\,\Ms$ of Paper~I.  The other three solid lines are the $F_{1P}(m_{1P,init})$ relations that correspond to the linear fits of Fig.~\ref{fig:degen} for $f_{red}=1.0, 0.6 {\rm ~and~} 0.3$ (same color coding).  Their expression is given by Eq.~6 in Paper~II, assuming a fixed bound fraction $F_{bound}^{VR}$ (i.e., $m_{th,init} = F_{bound}^{VR} m_{th}$ and $m_{1P,init} = F_{bound}^{VR} m_{1P,ecl}$).  The vertical dashed-dotted lines mark, for the same three reduction factors $f_{red}$, the thresholds $m_{th,init}$ from Fig.~\ref{fig:degen}.     

As already mentioned in Paper~II (its Sec.~5), slopes shallower than $\psi<1$ ($\psi=0.47\pm0.03$ when $f_{red}=1.0$ and $\psi=0.61\pm0.04 $ when $f_{red}=0.6$) can be interpreted in two different  ways.  In a first scenario, the stellar mass threshold for 2P-star formation is fixed but cluster pollution is not instantaneously completed, i.e., clusters keep forming 1P stars concurrently with 2P stars.  At the end of violent relaxation, this additional mass in 1P stars, $\Delta m_{1P}$, shows up as the horizontal arrows in Fig.~\ref{fig:ninst}: $\Delta m_{1P}$ adds to the threshold mass $m_{th,init}$ shown by the dashed-dotted vertical lines.  In a second scenario, clusters self-pollute instantaneously and the mass threshold for 2P-star formation depends on the pristine-star fraction, hence on the cluster mass.  At the end of violent relaxation, they are the solid lines labelled $m_{th,init}^{var}$ in Fig.~\ref{fig:ninst}.  $m_{th,init}$ is now the mass threshold for clusters with $F_{1P} \lesssim 1$ only.       
  
The case $f_{red}=0.3$ behaves differently.  Its $F_{1P}(m_{init})$ relation (pink track in Figs~\ref{fig:degen}) is slightly steeper than Paper~I model ($\psi=1.11\pm0.09>1.0$), which yields an initial mass in pristine stars that decreases with increasing cluster mass (since $m_{1P,init}$ decreases as $F_{1P}$ decreases; pink solid line in Fig.~\ref{fig:ninst}).  In this third scenario, more massive clusters experience therefore a lower mass theshold for 2P-star formation.  

Additional constraints are required to narrow the space of solutions obtained from sheer dynamical calculations.

\section{Summary and conclusions}
\label{sec:conclu}
It is four decades since \citet{cayrel86} 's original suggestion that Galactic halo \gcs formed two distinct and successive stellar generations, the first generation - a tiny fraction of the protoglobular cloud - providing the metal content and triggering the formation of the second one, whose stellar content was assumed to constitute today's globular clusters.  Cayrel 's proposal that \gcs are self-enriched systems was further developed or built on by \citet{dopita86}, \citet{morgan89}, \citet{bbt91,bbt95}, \citet{jehin98,jehin99}, \citet{par99,par00}, \citet{par01}, \citet{par04} and \citet{car10a}.  

The idea that Galactic halo \gcs synthetized their own metal content has since subdued, as they are now considered pre-enriched systems (i.e., their metal-content was part of the gas out of which they formed) rather than self-enriched systems.  Yet, the idea that \gcs have chemically evolved has taken root, eventually.  The riddle is no longer to explain their metal content, but, instead, to explain how they have enhanced some of their light elements at the expense of others, leading to their so-called pristine and polluted populations, the second one dominating the most massive globular clusters.  \\  

In this contribution, we have mapped the pristine-star fraction $F_{1P}$ of Galactic \gcs in dependence of their mass $m_{init}$ at secular-evolution onset, with the emphasis on the disk, Gaia-Enceladus and low-energy groups of \citet{mass19}.  We caution that cluster distributions in the $(m_{init},F_{1P})$ space remain necessarily incomplete as they include {\it surviving clusters only}.  We represent the $F_{1P}(m_{init})$ relations in a log-log space such that linear fits there get straightforwardly translated into power laws $F_{1P}= (m_{th,init}/m_{init})^\psi$ (Eq.~5 in Paper~II).  Figures~\ref{fig:disk} and \ref{fig:LEGE} show our results for the cluster dissolution time-scale $t_{diss}^{BM03}$ of \citet{bm03} and a time-scale $\simeq 3$-shorter, which corresponds to strongly primordially mass segregated clusters \citep{hag14}.    

We have highlighted the degeneracy between the normalization of the cluster dissolution time-scale and the slope and intercept of the $F_{1P}(m_{init})$ relation (see Fig.~\ref{fig:degen} for the disk clusters).  The shorter the cluster dissolution time-scale $t_{diss}=f_{red} t_{diss}^{BM03}$ (i.e., the smaller $f_{red}$), the greater the mass fraction lost by clusters over the past 12\,Gyr, the wider the gap between present-day and initial cluster masses, and the higher the mass threshold that clusters must have to host 2P stars at secular-evolution onset.  Because less massive clusters lose a greater fraction of their mass than high-mass clusters, shorter $f_{red}$ are also conducive to steeper $F_{1P}(m_{init})$.  

We have quantified the horizontal scatter $\Delta log_{10}(m_{init})$ of the $(m_{init},F_{1P})$ distributions as a function of $f_{red}$.  The disk group presents a tighter relation than the low-energy and Gaia-Enceladus groups (Fig.~\ref{fig:dhoriz}).  This could result from cluster-formation conditions that are more uniform in the early Galatic disk than in dwarf galaxies, or from disk protoglobular clumps achieving higher \stf efficiencies than clumps in dwarf galaxies (Fig.~\ref{fig:sfefb}).                   

An alternative explanation, however, is that the mass of the most inner and massive clusters is overestimated as a result of Eq.~\ref{eq:minit} neglecting dynamical friction.  This would artificially inflate the dispersion in initial mass of the low-energy and Gaia-Enceladus groups, whose census is dominated by inner clusters ($R_{eq}<3.1$\,kpc).  Because of their larger Galactocentric distance, disk clusters are largely immune to this effect.  

We have also investigated whether, for a given range of cluster initial masses, the fraction of polluted stars depends on metallicity (Sec.~\ref{sec:FeH}).  That does not seem to be the case for either of the three groups (Figs~\ref{fig:disk} and \ref{fig:LEGE}, where data points are color coded for cluster metallicity), but more accurate cluster initial masses and pristine-star fractions are required to answer this question definitively.  Specifically, a better handle on the degree of cluster primoridal mass segregation (i.e., hence on $f_{red}$), on the impacts of  dynamical friction (especially when $f_{red}<<1$) and of the time-variations in the Milky-Way gravitational potential is needed.  Pristine-star fractions $F_{1P}$ obtained from covering cluster areas that are representative of the whole clusters are needed as well.  
We also stress that the intuitive expectation of more metal-poor clusters hosting a greater fraction of 2P stars initially may not be so straightforward (our Fig.~\ref{fig:fehsketch} and Fig.~15 in \citealt{gieles25}).

Following a discussion of the uncertainties affecting photometric $F_{1P}$ estimates (Sec.~\ref{sec:phsp}), we have mapped what the slopes of the $F_{1P}(m_{init})$ linear fits imply for the initial mass in pristine stars $m_{1P,init}$.  Relations shallower than assumed in Paper~I ($\psi<1$) correspond to initial pristine-star masses that increase with the cluster total mass.  By contrast, relations steeper than Paper~I model ($\psi>1$) yield initial pristine-star masses that decrease with the cluster total mass (Fig.~\ref{fig:ninst}).  \\

Based on our results, a tantalizing question is whether the different groups of clusters, be they ex-situ or in-situ, share a universal $F_{1P}(m_{init})$ relation.  
Figure~\ref{fig:revobs} (its bottom panel) shows that, initially, the inner clusters may have been more massive than the outer clusters \citep[see also][]{frxburn09}.  Yet, we stress that such a comparison assumes that all clusters had similar degrees of primordial mass segregation and similar IMFs (i.e., the dissolution time-scale of all clusters is that of \citet{bm03} reduced by a similar factor $f_{red}=t_{diss}/t_{diss}^{BM03}$).  If different environments yield different degrees of primordial mass segregation and different IMFs, $f_{red}$ will differ from one cluster group to another, preventing us from safely concluding whether the clusters of one group are initially more massive than the clusters of another group.  

Data for the Small and Large Magellanic Clouds, whose respective cluster dissolution time-scales have been empirically estimated \citep{lam05a,par08}, suggest  distinct $F_{1P}(m_{init})$ relations.  Figure~6 in Paper~II shows that, for a given $F_{1P}$, clusters of the LMC are, already initially, more massive than their SMC siblings.  Paper~II suggests this initial shift to arise from LMC clusters achieving higher \stf efficiencies than SMC clusters, hence larger bound fractions $F_{bound}^{VR}$ after residual-gas expulsion.  It could also be the signature of a birth relation $F_{1P}(m_{ecl})$ that is environment dependent.  \\

\section*{acknowledgments} 

\small

G.P. acknowledges funding by the Deutsche Forschungsgemeinschaft (DFG: German Research Foundation) -- Project-ID 515414180.  She also thanks E.~Pancino for useful discussions.  G.P. is grateful to the Referee for a comprehensive and constructive report that has enriched the paper over its original version.  This research has made use of NASA 's Astrophysics Data System.


\bibliography{gparmentier_ApJ_2025a}{}
\bibliographystyle{aasjournal}



\end{document}